\documentclass[twocolumn]{aastex62}

\usepackage{graphicx}

\usepackage{amsmath,amssymb}

\newcommand{\hbeta}{H{$\beta$}}
\newcommand{\halpha}{H{$\alpha$}}

\newcommand{\OIII}{[O{\sevenrm\,III}]}

\newcommand{\OIIIb}{[O{\sevenrm\,III}]\,$\lambda$5007}
\newcommand{\OIIIc}{[O{\sevenrm\,III}]\,$\lambda\lambda$4959,5007}
\newcommand{\NII}{[N{\sevenrm\,II}]}

\newcommand{\NIIc}{[N{\sevenrm\,II}]\,$\lambda\lambda$6548,6584}
\newcommand{\SII}{[S{\sevenrm\,II}]}
\newcommand{\SIIa}{[S{\sevenrm\,II}]\,$\lambda$6717}
\newcommand{\SIIb}{[S{\sevenrm\,II}]\,$\lambda$6731}
\newcommand{\SIIab}{[S{\sevenrm\,II}]\,$\lambda\lambda$6717,6731}

\newcommand{\objfull}{SDSS~J084905.51$+$111447.2}
\newcommand{\obj}{SDSS~J0849+1114}
\newcommand{\hst}{{\it HST}}
\newcommand{\chandra}{{\it Chandra}}

 \font\sevenrm=cmr7 scaled 1000

\received{2019 September 24}
\revised{2019 November 2}
\accepted{2019 November 4}
\shorttitle{A Massive Black Hole Trio}
\shortauthors{Liu et al.}

\begin{document}

\title{A Trio of Massive Black Holes Caught in the Act of Merging\footnote{Based in part on observations made with the NASA/ESA {\it Hubble Space Telescope}. The observations were obtained at the Space Telescope Science Institute, which is operated by the Association of Universities for Research in Astronomy, Inc., under NASA contract NAS 5-26555. These observations are associated with program number GO 13112.}}

\email{xinliuxl@illinois.edu}

\author[0000-0003-0049-5210]{Xin Liu}
\affiliation{Department of Astronomy, University of Illinois at Urbana-Champaign, Urbana, IL 61801, USA}
\affiliation{National Center for Supercomputing Applications, University of Illinois at Urbana-Champaign, 605 East Springfield Avenue, Champaign, IL 61820, USA}

\author[0000-0001-9062-8309]{Meicun Hou}
\affiliation{School of Astronomy and Space Science, Nanjing University, Nanjing 210046, China}
\affiliation{Key Laboratory of Modern Astronomy and Astrophysics (Nanjing University), Ministry of Education, Nanjing 210046, China}
\affiliation{Department of Astronomy, University of Illinois at Urbana-Champaign, Urbana, IL 61801, USA}

\author[0000-0003-0355-6437]{Zhiyuan Li}
\affiliation{School of Astronomy and Space Science, Nanjing University, Nanjing 210046, China}
\affiliation{Key Laboratory of Modern Astronomy and Astrophysics (Nanjing University), Ministry of Education, Nanjing 210046, China}

\author[0000-0003-1991-370X]{Kristina Nyland}
\affiliation{National Research Council, resident at the U.S. Naval Research Laboratory, 4555 Overlook Avenue SW, Washington, DC 20375, USA}

\author[0000-0001-8416-7059]{Hengxiao Guo}
\affiliation{Department of Astronomy, University of Illinois at Urbana-Champaign, Urbana, IL 61801, USA}
\affiliation{National Center for Supercomputing Applications, University of Illinois at Urbana-Champaign, 605 East Springfield Avenue, Champaign, IL 61820, USA}

\author{Minzhi Kong}
\affiliation{Department of Physics, Hebei Normal University, No. 20 East of South 2nd Ring Road, Shijiazhuang 050024, China}
\affiliation{Department of Astronomy, University of Illinois at Urbana-Champaign, Urbana, IL 61801, USA}

\author[0000-0003-1659-7035]{Yue Shen}
\altaffiliation{Alfred P. Sloan Research Fellow}
\affiliation{Department of Astronomy, University of Illinois at Urbana-Champaign, Urbana, IL 61801, USA}
\affiliation{National Center for Supercomputing Applications, University of Illinois at Urbana-Champaign, 605 East Springfield Avenue, Champaign, IL 61820, USA}

\author[0000-0001-9720-7398]{Joan M. Wrobel}
\affiliation{National Radio Astronomy Observatory, P.O. Box 0, Socorro, NM 87801, USA}

\author[0000-0001-8492-892X]{Sijia Peng}
\affiliation{School of Astronomy and Space Science, Nanjing University, Nanjing 210046, China}
\affiliation{Key Laboratory of Modern Astronomy and Astrophysics (Nanjing University), Ministry of Education, Nanjing 210046, China}

\begin{abstract}
We report the discovery of \obj\ as the first known triple Type 2 Seyfert nucleus. It represents three active black holes that are identified from new spatially resolved optical slit spectroscopy using the Dual Imaging Spectrograph on the 3.5m telescope at the Apache Point Observatory. We also present new complementary observations including the {\it Hubble Space Telescope} Wide Field Camera 3 $U$- and $Y$-band imaging, {\it Chandra} Advanced CCD Imaging Spectrometer S-array X-ray 0.5--8 keV imaging spectroscopy, and NSF Karl G. Jansky Very Large Array radio 9.0 GHz imaging in its most extended A configuration. These comprehensive multiwavelength observations, when combined together, strongly suggest that all three nuclei are active galactic nuclei. While they are now still at kiloparsec-scale separations, where the host-galaxy gravitational potential dominates, the black holes may evolve into a bound triple system in $\lesssim$2 Gyr. These triple merger systems may explain the overly massive stellar cores that have been observed in some elliptical galaxies such as M87, which are expected to be unique gravitational wave sources. Similar systems may be more common in the early universe, when galaxy mergers are thought to have been more frequent.
\end{abstract}


\section{Introduction}\label{sec:intro}

Most local galaxies, if not all, are believed to harbor a massive black hole (BH) at the center\footnote{There are, however, galactic nuclei that do not appear to contain massive black holes (MBHs), such as M33 \citep{Gebhardt2001}.} \citep{KormendyHo2013}. Black hole pairs and triples are natural outcomes of the hierarchical galaxy and BH assembly process. Studies based on the Sloan Digital Sky Survey \citep[SDSS;][]{York2000} galaxies show that ${\lesssim}2$\% of the (major) merging systems involve three galaxies \citep[e.g.,][]{Darg2010}. While hundreds of active BH pairs (i.e., dual active galactic nuclei) and candidates have been found in galaxy mergers \citep[e.g.,][]{Liu2011a}, only three kiloparsec-scale (kpc-scale for short) triple-AGN candidates are known \citep{barth08,Liu2011,schawinski11}. However, their identification based on optical diagnostic emission-line ratios has been inconclusive. Alternative ionizing sources remain possible, such as stellar and shock-heating processes.

\begin{figure*}
\centerline{
\includegraphics[width=1.0\textwidth]{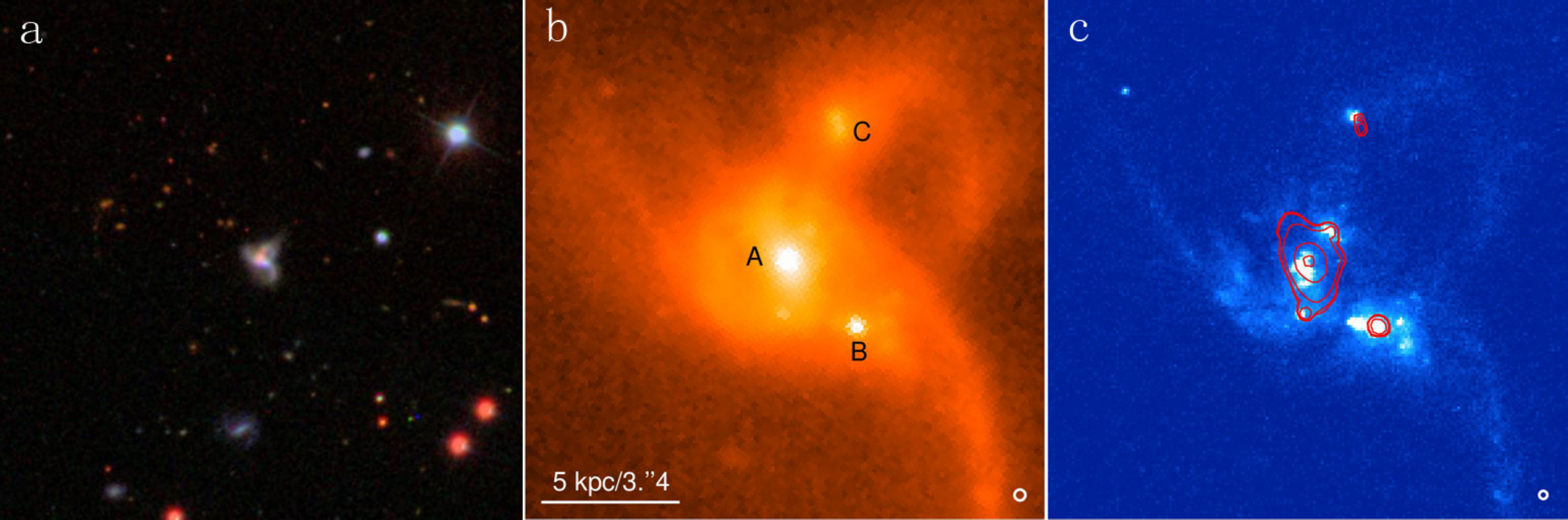}
}
\caption{Optical and near-IR images of \obj . 
Panel (a) shows SDSS $gri$-color composite image of a 200''$\times$200'' (corresponding to $\sim$300 kpc$\times$300 kpc) field-of-view centering on the galaxy. The galaxy is a relatively open field with no large companion within a $\sim$150 kpc radius. Panels (b) and (c) show zoom-in images of the galaxy taken with \hst /WFC3 in IR/F105W ($Y$) and UVIS/F336W ($U$) bands. Circles at the bottom indicate the \hst\ PSF sizes. Contours are in the \hst\ $Y$ band. North is up and east is to the left.}
\label{fig:optical_image}
\end{figure*}

The galaxy \objfull\ (hereafter \obj\ for short) at redshift $z=0.078$ contains three optical stellar nuclei within a projected $\sim$5 kpc radius that show a disturbed morphology (Figure \ref{fig:optical_image}). \obj\ was identified as a candidate to host a kpc-scale triple MBH from the largest sample of optically selected candidate AGN pairs \citep{Liu2011a} based on the SDSS Seventh Data Release \citep{SDSSDR7}. The SDSS has two spectra taken with a 3$''$ diameter fiber centered on Nucleus B (projected at 2\farcs3 or 3.4 kpc to the southwest of A) and C (projected at 3\farcs6 or 5.2 kpc to the northwest of A), classifying B as a Type 2 Seyfert and C as a low-ionization nuclear
emission-line region\footnote{C is classified as a LINER by \citet{Liu2011a} based on emission-line ratios measured from the SDSS spectrum because the \citet{ho97} criterion is adopted to separate Seyferts from LINERs; it would have been classified as a Seyfert if the \citet{Schawinski2007} criterion were adopted. Our new APO/DIS measurements unambiguously classify C as a Seyfert regardless of which criterion is used.} \citep[LINER;][]{Liu2011a}. 

\begin{figure*}
  \centering
    \includegraphics[width=0.8\textwidth]{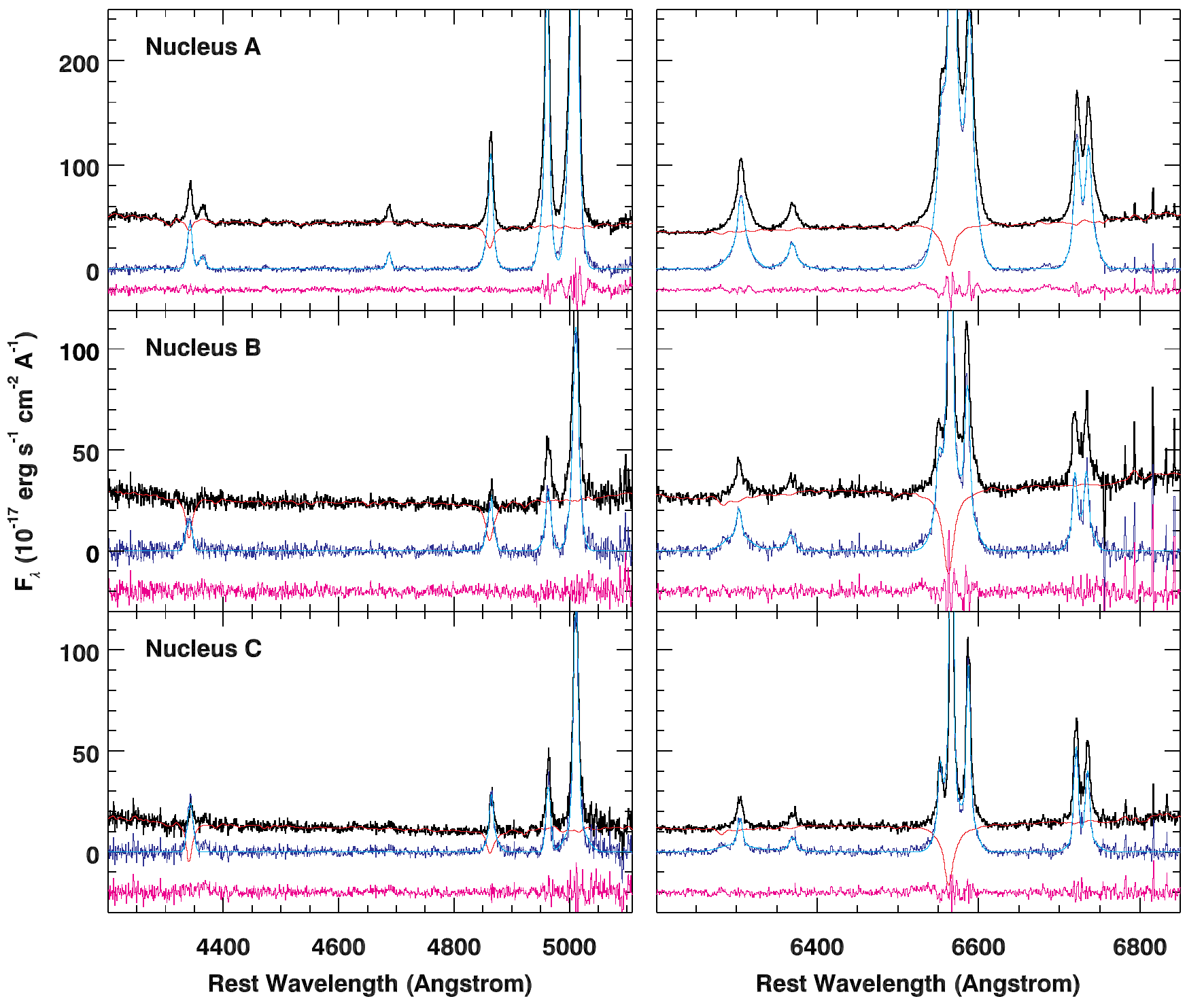}
    \caption{APO 3.5m/DIS optical long-slit spectra show three emission-line nuclei in \obj . 
    Data are shown in black, and the best-fit stellar continuum model is shown in red. Also shown are the continuum-subtracted spectrum (blue), the best-fit emission-line model (cyan), and the residual (magenta), offset vertically for clarity.
    }
    \label{fig:spec}
\end{figure*}

\begin{figure}
  \centering
    \includegraphics[width=0.5\textwidth]{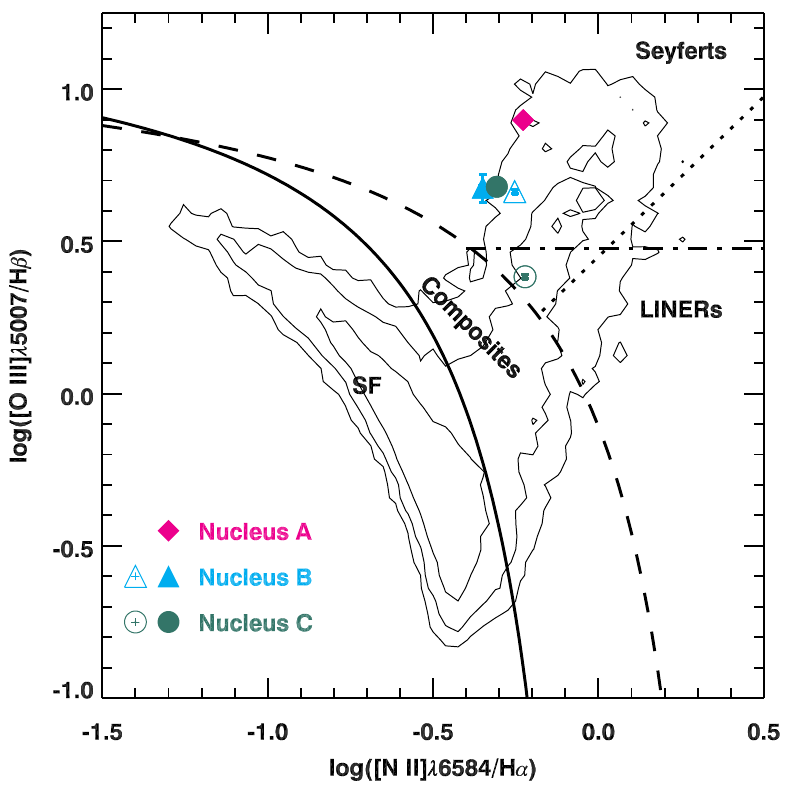}
    \caption{Optical diagnostic emission-line ratio diagram \citep{bpt,veilleux87} classifies all three nuclei in \obj\ as Type 2 Seyferts. 
    Filled symbols denote measurements based on our new APO/DIS slit spectra, whereas open symbols represent those from SDSS fiber spectra. Contours show number densities of SDSS emission-line galaxies \citep{kauffmann03}. Curves and lines show the empirical separation between H {\tiny II} regions and AGN \citep[][solid]{kauffmann03}, the theoretical ``starburst limit'' \citep[][dashed]{kewley01}, and the empirical divisions between Seyferts and LINERs (\citealt{Schawinski2007}, dotted, and \citealt{ho97}, dash-dotted).
    }
    \label{fig:bpt}
\end{figure}

To understand the physical nature of Nucleus A and to more clearly separate light coming from each individual nucleus, we here present spatially resolved follow-up optical spectroscopy with the Dual Imaging Spectrograph (DIS) on the 3.5m telescope at the Apache Point Observatory. After carefully separating the light from each nucleus and subtracting the host-galaxy stellar continuum (Figure \ref{fig:spec}), all three nuclei are optically classified as Type 2 (obscured) Seyferts based on the diagnostic narrow emission-line ratios according to the classical BPT diagram (Figure \ref{fig:bpt}). 

\obj\ represents the first known case of a triple Type 2 Seyfert nucleus. Unlike \obj , previous triple-AGN candidates all contain at least one nucleus classified as either LINER or AGN-H {\tiny II} composite, which may be due to excitation by evolved stars, star formation, and/or shock heating instead of AGN ionization. For example, the disk galaxy NGC 3341 contains three nuclei that are optically classified as a Seyfert, a LINER, and a LINER or LINER-H {\tiny II} composite \citep{barth08}. X-ray and radio follow-up observations suggest that both the primary galaxy and one of the secondary dwarf companions are unlikely to host an AGN \citep{Bianchi2013}. Another potential candidate was reported at $z=1.35$ based on {\it Hubble Space Telescope} ({\it HST}) grism spectroscopy \citep{schawinski11}, although the spectral resolution was too low to reliably resolve the individual components, which may be from a clumpy star-forming disk galaxy rather than a triple AGN. Yet another candidate, now disputed, J1502+1115 (at $z=0.39$), was identified by the Very Long Baseline Interferometry \citep[VLBI;][]{Deane2014}, although its two nuclei were later found to be radio hot spots instead of a pair of BHs \citep{Wrobel2014}. Finally, in another candidate from the SDSS candidate AGN pair sample, SDSSJ1027+1749, only one of the nuclei is optically classified as a Seyfert, whereas the other two are a LINER and a composite \citep{Liu2011}.

While the triple Type 2 Seyfert nucleus suggests three active MBHs in \obj , optical diagnostics may be inconclusive. There could be only one or two active BHs ionizing all three merging components, resulting in three Seyfert-like nuclei, although this scenario is unlikely considering their positions on the BPT diagram. To further determine the excitation mechanism of the triple Type 2 Seyfert nucleus in \obj , here we also present a comprehensive follow-up campaign including {\it HST} Wide Field Camera 3 (WFC3) $U$- and $Y$-band imaging, {\it Chandra} Advanced CCD Imaging Spectrometer S-array (ACIS-S) X-ray 0.5--8 keV imaging spectroscopy, and NSF Karl G. Jansky Very Large Array (VLA) radio 9.0 GHz imaging in its most extended A configuration. 

The paper is organized as follows. \S \ref{sec:obs} presents our target selection and new observations from APO 3.5m/DIS (\S\S \ref{subsec:apo}, \ref{subsec:opt_spec}), HST/WFC3 (\S\S \ref{subsec:hst}, \ref{subsec:opt_img}), \chandra\ ACIS-S (\S \ref{subsec:chandra}), and VLA A-config (\S \ref{subsec:vla}). \S \ref{sec:result} shows our results on the host-galaxy internal dust extinction (\S \ref{subsec:dust}), stellar mass (\S \ref{subsec:stellarmass}) and nuclear star formation rate (SFR; \S \ref{subsec:sfr}), X-ray contribution from nuclear star formation (\S \ref{subsec:sf_xray}), photoionization diagnostics based on optical spectra (\S \ref{subsec:photo}), and the nature of the nuclear ionizing sources (\S \ref{subsec:nature}). We then discuss the implications of our results in \S \ref{sec:discuss}. Finally, we summarize our main findings and discuss future work in \S \ref{sec:sum}. Throughout this paper, we assume a concordance cosmology with $\Omega_m = 0.3$, $\Omega_{\Lambda} = 0.7$, and $H_{0}=70$ km s$^{-1}$ Mpc$^{-1}$ and use the AB magnitude system \citep{oke74}.

\section{Observations, Data Reduction, and Data Analysis}\label{sec:obs}

\subsection{Target Selection}\label{subsec:target_selection}

%\newgeometry{margin=1cm}
%\begin{landscape}
\begin{table*}
\caption{Host-galaxy Properties of \obj\ }
%\centering
\begin{tabular}{ccccccccccccc}
  \hline
  \hline
& & & $r$ & $z$ & $u - z$ & $M_{z}$ & $M_{z,\,{\rm c}}$ & $m_Y$ & $m_{U}$ & log$M_{\ast}$ & log$M_{\ast,\,{\rm ap}}$ & log$M_{\bullet}$ \\
ID & SDSS Designation & Redshift & (mag) & (mag) & (mag) & (mag) & (mag) & (mag) & (mag) & ($M_{\odot}$) & ($M_{\odot}$) & ($M_{\odot}$)  \\
(1) & (2) & (3) & (4) & (5) & (6) & (7) & (8) & (9) & (10) & (11) & (12) & (13) \\
\hline
A & J084905.51+111447.2\dotfill & 0.0775 & 15.90 & 15.21 & 2.38 & $-$22.52 & $-$23.26 & 16.62 & 19.72 & 11.3 & 10.6 & 7.5 \\
B & J084905.39+111445.8\dotfill & 0.0778 & 18.36 & 18.39 & 1.09 & $-$19.35 & $-$21.34 & 17.83 & 19.26 & 10.0 & 10.1 & 6.4 \\
C & J084905.42+111451.0\dotfill & 0.0779 & 17.49 & 16.89 & 2.79 & $-$20.26 & $-$21.06 & 18.40 & 21.42 & 10.5 & 9.7 & 6.7 \\
\hline
\end{tabular}
\tablecomments{
Column 1: nucleus ID as labeled in Figure \ref{fig:optical_image}. 
Column 2: SDSS names with J2000 coordinates given in the form of ``hhmmss.ss+ddmmss.s''. 
Column 3: redshift measured from stellar continuum and emission-line fitting. 
Columns 4 and 5: SDSS $r$- and $z$-band model magnitude. 
Column 6: color calculated from SDSS $u$- and $z$-band model magnitudes.
Columns 7 and 8: SDSS k-corrected $z$-band absolute magnitude before and after correction for internal dust extinction. 
Columns 9 and 10: \hst\ Y- and U-band magnitudes from aperture photometry using a 1$''$ radius circle centered on each nucleus. 
Column 11: total host galaxy stellar mass estimate derived from Equation \ref{eq:stellar_mass}. 
Column 12: stellar mass estimate within the central 1$''$ radius circle of each nucleus. 
Column 13: BH mass estimate inferred from host galaxy total stellar mass assuming the $M_{\bullet}$--$M_{\ast}$ relation observed in local broad-line AGN with virial BH mass estimates \citep{Reines2015}.
}
\label{tab:host}
\end{table*}
%\end{landscape}
%\restoregeometry
%%
%%

\obj\ (Table \ref{tab:host}) was selected from an \OIII -selected AGN pair sample \citep{Liu2011a} systematically identified in the SDSS DR7 \citep{SDSSDR7}. We searched for kpc-scale triple-AGN candidates among kpc-scale AGN pairs with a companion galaxy within 10 kpc in projection from the first two nuclei. We then conducted spatially resolved optical spectroscopy to measure redshifts and to estimate excitation mechanisms for all nuclei. While in most systems the third nucleus turned out to be an inactive neighbor, we found three candidate kpc-scale triple AGNs \citep{Liu2011} for further study. \obj\ is unique among the three candidates in that it is the only one with a triple Type 2 Seyfert nucleus. \obj\ was also independently identified as a dual AGN/triple-AGN candidate \citep{Pfeifle2019} in a sample of IR-selected mergers based on the {\it Wide-field Infrared Survey Explorer} data \citep{Wright2010}. After we submitted this manuscript, another work \citep{Pfeifle2019a} also independently proposed \obj\ as a triple AGN, although we argue that the comprehensive evidence shown in the present paper (in particular the stringent constraints on the alternative scenario of a nuclear starburst based on the high-resolution VLA and {\it HST} $U$-band imaging) is significantly stronger.   

We distinguish between kpc-scale triple AGN \citep[][i.e., the focus of this work]{Liu2011} from physical triple quasars \citep{djorgovski07,Farina2013,Hennawi2015}. The former have typical projected physical separations of a few kiloparsec and velocity offsets of $<$300 km s$^{-1}$ when the host galaxies are already in direct gravitational interactions. The latter have typical projected physical separations of a few hundred kiloparsec and velocity offsets of $<$1000 km s$^{-1}$, suggesting that the quasars belong to the same dark matter halo, although it is unclear if their host galaxies are already in direct gravitational interaction with each other. 

\subsection{APO 3.5m/DIS Long-slit Spectroscopy}\label{subsec:apo}

We obtained long-slit spectra for \obj\ on the night of 2011 March 6 UT using the Dual Imaging Spectrograph (DIS) on the Apache Point Observatory 3.5 m telescope. The sky was non-photometric, and the seeing ranged from 0\farcs8 to 2\farcs0, with a median of $\sim$1\farcs2. The field of view of DIS was 4$'\times$6$'$ with a pixel size of 0\farcs414. We used a 1\farcs5$\times$6$'$ slit with the B1200+R1200 gratings centered at 510 and 700 nm. The spectral coverage was 450--560 nm and 640--760nm with an instrumental resolution of $\sigma_{{\rm inst}} \approx$60 and 30 km s$^{-1}$ and a dispersion of 0.62 and 0.58 {\AA} pixel$^{-1}$ in the blue and red channels. The slit was oriented at a position angle of 43$^{\circ}$ and 159$^{\circ}$ to go through Nucleus A and B, and A and C (Figure \ref{fig:xray_radio}). The total effective exposure time was 5400 s, 2700 s, and 2700 s for A, B, and C. We observed standard stars G191B2B and HZ44 to calibrate spectrophotometry. 

We reduced the DIS data following standard IRAF\footnote{IRAF is distributed by the National Optical Astronomy Observatory, which is operated by the Association of Universities for Research in Astronomy (AURA) under cooperative agreement with the National Science Foundation.} procedures \citep{tody86}. We extracted 1D spectra using a 2$''$ diameter aperture for each nucleus. We applied a telluric correction from standard stars to the extracted 1D spectra. The median signal-to-noise ratio (S/N) we achieved is $\sim$10--40 pixel$^{-1}$ (Figure \ref{fig:spec}).

\subsection{Optical Spectral Analysis}\label{subsec:opt_spec}

We first measure the host-galaxy stellar continuum in the emission-line free regions of the observed spectrum and then model the emission-line flux over the host-subtracted spectrum. For the host-galaxy spectral fitting, we use the penalized pixel-fitting (pPXF) method \citep{Cappellari2004}, which works directly in the pixel space using the maximum penalized likelihood formalism to extract the most information from the spectra while suppressing noise in the solution. For each nucleus, we measured the host redshift by fitting the continuum with galaxy templates produced by population synthesis models \citep{bc03} following the procedure of \citet{Liu2009}. We convolved the galaxy templates with the stellar velocity dispersion $\sigma_{\ast}$ measured over small spectral windows containing strong stellar absorption features.

After subtracting the stellar continuum using the best-fit pPXF model, we fit the narrow emission lines simultaneously with multiple Gaussian models constrained to have the same velocity and line width using the spectral fitting code qsofit \citep{Guo2018,Shen2019}. We also verified the host redshift by measuring the velocity of the combined set of narrow emission lines. 

Figure \ref{fig:spec} shows the rest-frame 1D spectrum and our best-fit models for the stellar continuum and the continuum-subtracted emission lines. Table \ref{tab:line} lists the emission-line measurements from our best-fit models.

%%%%%%%%%%%%%%%%%%%%%%%%%%%%%%%%%%%%%%%%%%%%%%%%%%%%%%%%%%%%%%%%
%\newgeometry{margin=1cm}
%\begin{landscape}
\begin{table*}
\caption{Emission-line properties of \obj\ } 
%\centering
\begin{tabular}{ccccccccccccc}
  \hline
  \hline
& & & & & & & log $n_e$ & & & $\sigma_{{\rm gas}}$ & $\sigma_{\ast}$ & log$M_{\bullet}$ \\
ID & $\frac{{\rm [O\,III]}}{{\rm H}\beta}$ & $\frac{{\rm [N\,II]}}{{\rm H}\alpha}$ & $\frac{{\rm [O\,I]}}{{\rm H}\alpha}$ & $\frac{{\rm [S\,II]}}{{\rm H}\alpha}$ & $\frac{{\rm H}\alpha}{{\rm H}\beta}$ & $\frac{{\rm [S\,II]}\lambda6717}{{\rm [S\,II]}\lambda6731}$ & (cm$^{-3}$) & $\frac{{\rm [O\,III]}\lambda\lambda 4959,5007}{{\rm [N\,II]}\lambda\lambda6548,6584}$ & $\frac{{\rm [O\,III]}\lambda\lambda 4959,5007}{{\rm [S\,II]}\lambda\lambda6717,6731}$ & (km s$^{-1}$) & (km s$^{-1}$) & ($M_{\odot}$)  \\
(1) & (2) & (3) & (4) & (5) & (6) & (7) & (8) & (9) & (10) & (11) & (12) & (13) \\
\hline
A$^{\dagger}$ & 7.92 & 0.592 & 0.200 & 0.445 & 4.75 & 1.15 & 2.49 & 6.87 & 13.06 & 134$\pm$3 & 193$\pm$19 & 7.9 \\
B$^{\dagger}$ & 4.75 & 0.447 & 0.179  & 0.351 & 6.45 & 1.01 & 2.75 & 6.53 & 12.47 & 101$\pm$22 & 119$\pm$17 & 7.1 \\
B$^{\ddagger}$ & 4.61 & 0.558 & 0.108  & 0.427 & 4.21 & 1.29 & 2.15 & 3.83 & 7.02 & 162$\pm$1 & 101$\pm$8 & 6.8 \\
C$^{\dagger}$ & 4.78 & 0.493 & 0.140 & 0.442 & 5.00 & 1.38 & 1.77 & 4.66 & 7.50 & 113$\pm$14 & 144$\pm$53 & 7.4 \\
C$^{\ddagger}$ & 2.42 & 0.600 & 0.108 & 0.479 & 4.89 & 1.05 & 2.68 & 2.14 & 3.84 & 162$\pm$1 & 109$\pm$17 & 6.9 \\
\hline
\end{tabular}
\tablecomments{
Column 1: same as Column 1 in Table \ref{tab:host}. The dagger indicates our own measurements from DIS slit spectra, and the double cross denotes those from the MPA-JHU SDSS DR7 value added catalog based on SDSS fiber spectra. 
Columns 2--7, 9, and 10: emission-line intensity ratio. \OIII /\NII\ and \OIII /\SII\ have been corrected for dust reddening using the Balmer decrement method. 
Column 8: gas electron density inferred from the emission-line intensity ratio \SIIa /\SIIb . 
Column 11: gas velocity dispersion from modeling the continuum-subtracted emission lines. 
Column 12: stellar velocity dispersion estimate from modeling the stellar continuum. 
Column 13: BH mass estimate inferred from host-galaxy stellar velocity dispersion assuming the $M_{\bullet}$--$\sigma_{\ast}$ relation in Equation \ref{eq:bh_vdisp}.
}
\label{tab:line}
\end{table*}
%\end{landscape}
%\restoregeometry
%%
%%

\subsection{HST WFC3 $U$- and $Y$-band Imaging}\label{subsec:hst}

We observed \obj\ using the WFC3 on board the \hst\ on 2013 March 18 UT (Program GO 13112; PI Liu). Images were taken in the UVIS/F336W ($U$ band, with pivot $\lambda_{{\rm p}}=$335.5 nm and width of 51.1 nm; \citealt{dressel10}) and IR/F105W (wide $Y$ band, with $\lambda_{{\rm p}}=$1055.2 nm and width of 265.0 nm, \citealt{dressel10}) filters in a single orbit. The total net exposure times were 2151 s in $U$ and 239 s in $Y$. To cover \obj\ and the nearby field adequately for background subtraction, we adopted a 1k$\times$1k subarray in $U$ for a field of view (FOV) of 40$''\times$37$''$ and a 512$\times$512 subarray in $Y$ for an FOV of 72$''\times$64$''$. The WFC3 UVIS CCD has a pixel size of 0\farcs039, and the IR detector has a pixel size of 0\farcs13. We dithered the observations to properly sample the point-spread functions (PSFs) while rejecting cosmic rays and bad pixels.   

We reduced the \hst\ data following standard procedures. Geometric distortion and pixel area effects were corrected for. We combined dithered frames while rejecting cosmic rays and hot pixels. The final image has a pixel scale of 0\farcs06 for the $Y$-band image to be Nyquist-sampled. Typical relative astrometric accuracy is 0\farcs004 for UVIS and 0\farcs01 for IR images. The absolute astrometric accuracy of the reduced WFC3 image is limited by the positional uncertainty of guide stars and the calibration uncertainty of the fine guidance sensor to the instrument aperture. For better absolute astrometric accuracy, we registered the $Y$-band image to the SDSS astrometry \citep{pier03} using reference objects selected from the FOV with measured SDSS astrometry. The absolute astrometric uncertainty of the registered image is estimated as $\sim$0\farcs15, which includes $\sim$0\farcs055, the typical SDSS astrometric uncertainty \citep{pier03}, and $\sim$0\farcs085, the statistical uncertainty of the standard deviation of the WCS fit. We tie the $U$-band astrometry to the $Y$ band because of the small relative astrometric uncertainty between adjacent exposures. Figure \ref{fig:optical_image} shows the reduced and calibrated \hst\ images for \obj .

\begin{figure*}
  \centering
    \includegraphics[width=1.0\textwidth]{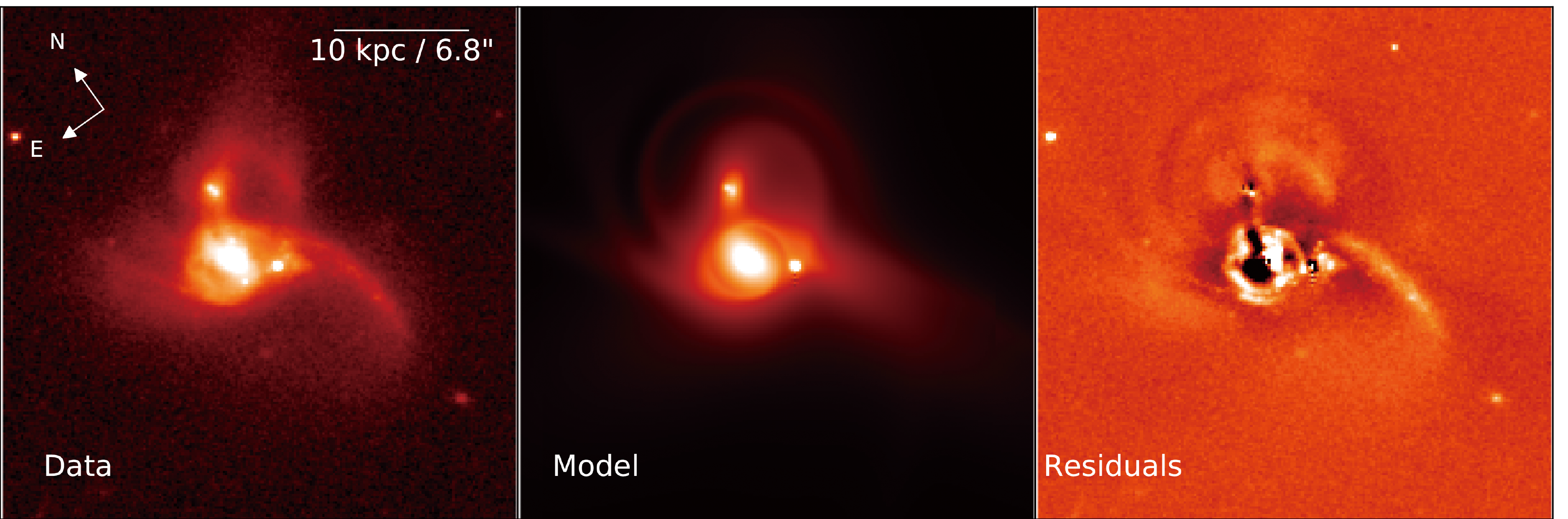}
    \caption{
   Host-galaxy structural decomposition. 
   Shown are HST WFC3 $Y$ band image, our best-fit model from GALFIT analysis, and the fitting residuals (i.e., data$-$model). Table \ref{tab:galfit} lists the best-fit model parameters. 
    }
    \label{fig:galfit}
\end{figure*}

\begin{figure}
  \centering
    \includegraphics[width=0.5\textwidth]{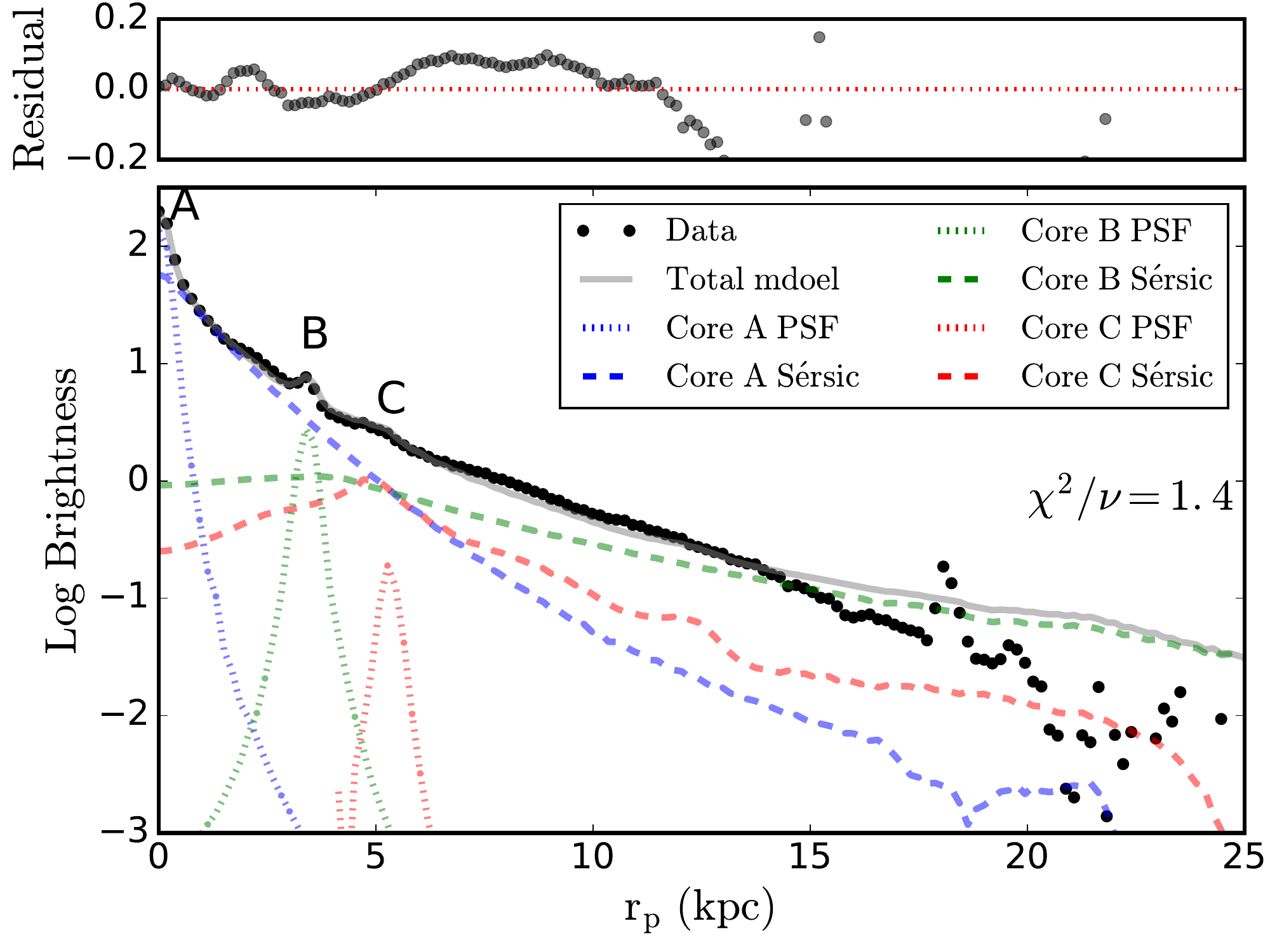}
    \caption{
    {\it HST} $Y$-band radial surface brightness profile.
    For each merging component, we use one S${\rm \acute{e}}$rsic model for a disk or a resolved bulge component and a PSF model for any unresolved component (e.g., from a compact stellar bulge, a nuclear starburst, or continuum emission from an unresolved narrow line region).
    Our best-fit model for Nucleus A is dominated by a disk component with S${\rm \acute{e}}$rsic index $n=1.3$. Nucleus B is dominated by an extended disk component with S${\rm \acute{e}}$rsic index $n=1.5$. Nucleus C is dominated by a resolved bulge with S${\rm \acute{e}}$rsic index $n=3.5$, although it does show an extended disk component in the residual image that likely is being tidally striped by A.   
    }
    \label{fig:radial_profile}
\end{figure}

\subsection{Analysis of Host-galaxy Morphology and Structural Decomposition}\label{subsec:opt_img} 

We use the \hst\ $Y$-band image to study host-galaxy morphology and old stellar populations. Given the highly disturbed morphology in the complex merging galaxy, we use GALFIT \citep{peng10} to model the 2D surface brightness profile with multiple structural components. Instead of achieving a perfect fit to the data, our goal is to try and decompose the three galaxies from each other and to try and decompose the disk and bulge components, if any, in each galaxy. We use stars within the FOV to model the PSF and employ a constant sky background in the model. For each galaxy, we use one S${\rm \acute{e}}$rsic model for a disk or a resolved bulge component and a PSF model for any unresolved component (e.g., from a compact stellar bulge, a nuclear starburst, or continuum emission from an unresolved narrow line region). The S${\rm \acute{e}}$rsic model profile is given by
\begin{equation}\label{eq:sersic}
\Sigma(r) = \Sigma_e \, {\rm exp} \bigg [ - \kappa \bigg( \Big ( \frac{r}{r_e} \Big )^{1/n} -1  \bigg ) \bigg ]
\end{equation}
where $\Sigma(r)$ is the pixel surface brightness at radial distance $r$, $\Sigma_e$ is the pixel surface brightness at the effective radius $r_e$, and $\kappa$ is a parameter related to the S${\rm \acute{e}}$rsic index $n$. $n=1$ for an exponential profile, while $n=4$ (corresponding to $\kappa=7.67$) for a de Vaucouleurs profile. Bulge-dominated galaxies have high $n$ values (e.g., $n>2$), while disk-dominated galaxies have $n$ close to unity. We also tried using an additional S${\rm \acute{e}}$rsic for any resolved bulge, but adding another component did not significantly improve the fit. 

Figure \ref{fig:galfit} shows our best-fit GALFIT model. The residual image still shows significant patterns, which is expected in highly disturbed mergers. Figure \ref{fig:radial_profile} shows the radial surface brightness profiles for individual components. Table \ref{tab:galfit} lists the best-fit model parameters. Our best-fit model for Nucleus A is dominated by a disk component with S${\rm \acute{e}}$rsic index $n=1.3$. Nucleus B is dominated by an extended disk component with S${\rm \acute{e}}$rsic index $n=1.5$, which is likely being tidally striped by A. Nucleus C is dominated by a bulge component with S${\rm \acute{e}}$rsic index $n=3.5$, although it does show an extended disk component in the residual image, which is again likely being tidally striped by A.

\begin{table}
\begin{center}
\caption{HST $Y$-band Photometric Decomposition Results from GALFIT Analysis} 
\begin{tabular}{cccccc}
  \hline
  \hline
 & m$_{Y,\,{\rm PSF}}$ & m$_{Y,\,{\rm S\acute{e}rsic}}$ & R$_{\rm e}$ &  & \\
ID & (mag) & (mag) & (kpc) & n & B/T \\
(1) & (2) & (3) & (4) & (5) & (6) \\
\hline
A & 18.4 & 15.7 & 2.4 & 1.3 & 0.076 \\
B & 18.9 & 16.1 & 8.2 & 1.5 & 0.071 \\
C & 21.4 & 17.0 & 3.9 & 3.5 & - \\
\hline
\end{tabular}
\end{center}
\tablecomments{
Column 1: same as Column 1 in Table \ref{tab:host}. 
Column 2: $Y$-band magnitude of the PSF component in our best-fit GALFIT model. 
Column 3: $Y$-band magnitude of the S${\rm \acute{e}}$rsic component in our best-fit GALFIT model. 
Column 4: effective radius for the best-fit S${\rm \acute{e}}$rsic component. 
Column 5: best-fit S${\rm \acute{e}}$rsic index. 
Column 6: $Y$-band bulge-to-total luminosity ratio estimated using the best-fit GALFIT model. We failed to measure this for Nucleus C, which does have an extended disk component, but is being tidally striped by the primary galaxy.
}
\label{tab:galfit}
\end{table}
%\end{landscape}
%%
%%

\subsection{Chandra ACIS-S X-Ray 0.5--8 keV Imaging Spectroscopy, X-Ray Data Reduction, and Data Analysis}\label{subsec:chandra}

We observed \obj\ with the ACIS-S on board the \chandra\ X-ray Observatory on 2013 February 8 UT (Program GO3-14104A, ObsID=14969; PI Liu). \obj\ was observed on-axis on the S3 chip with an exposure time of 20 ks. Following our Cycle 14 observation, there was a Cycle 17 ACIS-S observation, taken on 2016 March 3 UT with an exposure time of 21 ks (ObsID=18196; PI Satyapal). We combine both observations to maximize the S/N. The total effective exposure time of the combined image is 41 ks.

\begin{figure*}
  \centering
    \includegraphics[width=1.0\textwidth]{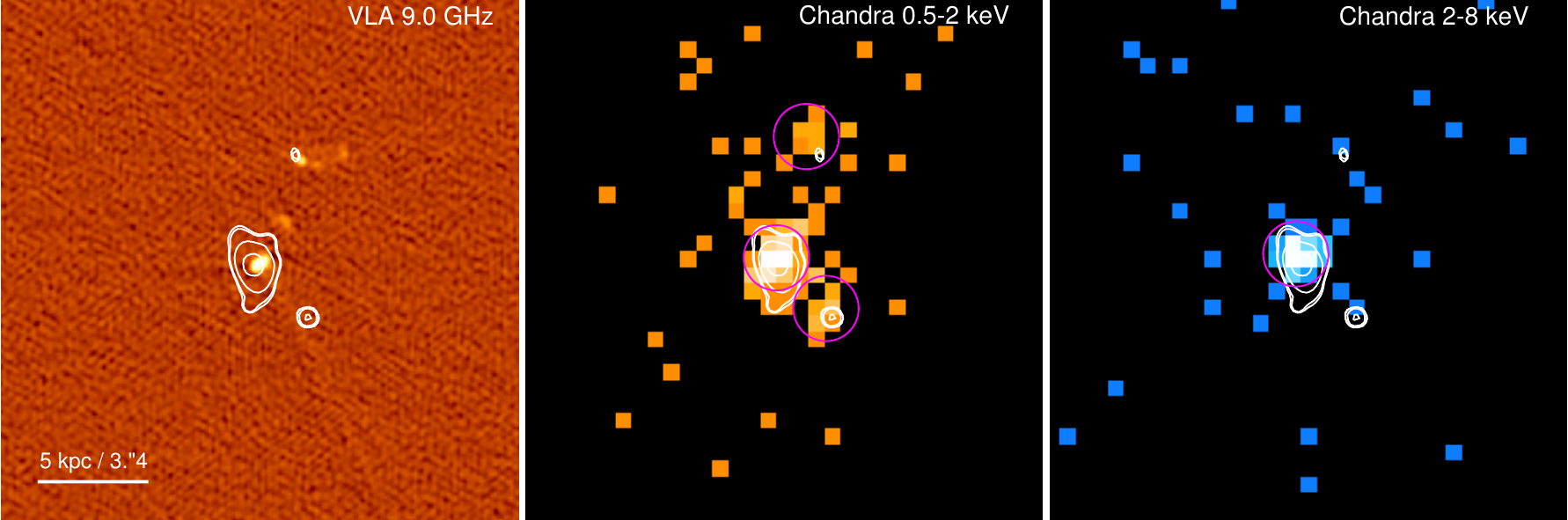}
    \caption{VLA radio and \chandra\ X-ray images for \obj .
    Left: 1-hr VLA 9.0 GHz image (beam size $\sim$0\farcs19$\times$0\farcs19). Middle and right: 41-ks \chandra\ ACIS-S images (unsmoothed) in soft and hard bands. Magenta circles are 1$''$  in radius, representing the extraction apertures in X-ray measurements.  
    White contours are in the \hst\ $Y$ band. North is up and east is to the left.
    }
    \label{fig:xray_radio}
\end{figure*}

We reprocessed the \chandra\ data using CIAO v4.8 with the calibration files CALDB v4.7.0 following standard procedures. The light curve of each observation was examined to ensure that there was no significant particle background. We generate counts maps, exposure maps and PSF maps at the original pixel scale in all bands (0\farcs492 pixel$^{-1}$). We weigh exposure maps with a fiducial model for Type 2 Seyferts \citep{green09} assuming an absorbed power law with a photon-index $\Gamma_{{\rm X}}$=1.7 and a column density $N_{{\rm H}}=$10$^{22}$ cm$^{-2}$. The CIAO tool {\it wavdetect} is used to search for tentative X-ray sources coincident with the three nuclei. For each resultant source, we derive a background-subtracted and exposure-map--corrected photon flux from a 1$''$ radius circle (representing $\sim$ 90\% of the enclosed energy) using the CIAO tool {\it aprate}. This extraction aperture ensures that there is no overlap between different nuclei. The local background is extracted typically between 2$''$ and 5$''$, and other detected point sources in this region are removed. Nuclei with 3$\sigma$ photon flux lower limits greater than 0 are considered as a firm detection. For the non-detected nuclei, we estimate their 3$\sigma$ photon flux upper limits. Nucleus A is detected in both soft and hard bands, whereas B and C are detected in the soft band only. Figure \ref{fig:xray_radio} shows the ACIS image of \obj\ in the soft (0.5--2 keV) and hard (2--8 keV) bands. Given the low count level, we do not apply any smoothing to avoid artifacts. All X-ray detections are consistent with the \hst\ $Y$-band nucleus positions within astrometric uncertainties.

%\newgeometry{margin=1cm}
%\begin{landscape}
%\end{landscape}
%\restoregeometry

\begin{figure}
  \centering
    \includegraphics[width=0.35\textwidth,angle=270]{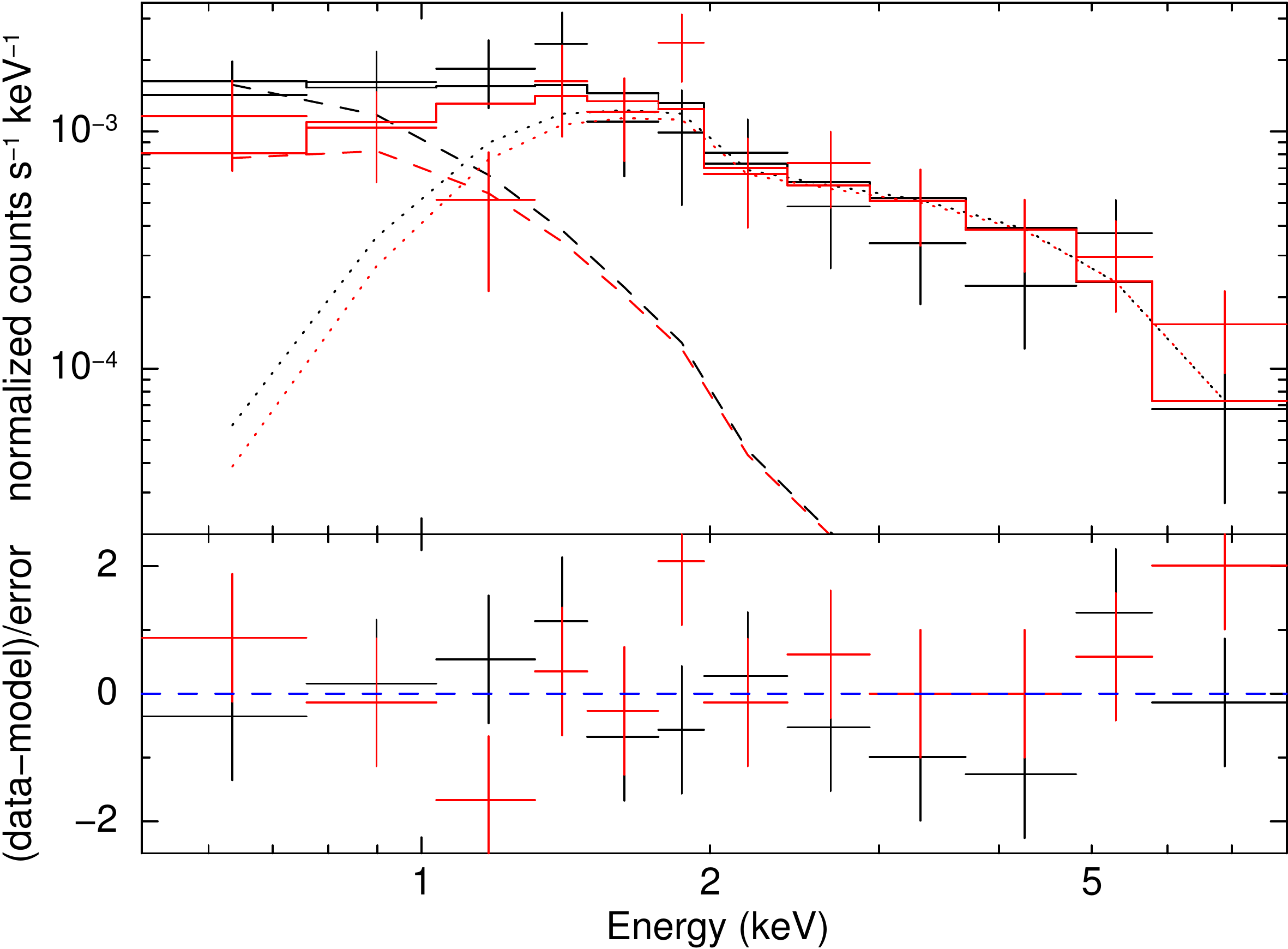}
    \caption{
    X-ray spectral modeling for Nucleus A.
    The upper panel shows the data and the best-fit model, and the lower panel shows the fitting residual normalized by the error. The earlier 2013 observation is shown in black, and the later observation in 2016 is shown in red. The small apparent difference in the two folded spectra at energies below $\sim$2 keV is due to the degrading response of ACIS with time. The actual fit is performed on the unbinned spectra, while we show here for illustration the spectra adaptively binned to achieve a combined S/N $\geq$ 3. The best-fit model (solid curves) consists of an absorbed power law (dotted curves) for the AGN and a power law  (dash curves) for circumnucleus emission from X-ray binaries and/or hot gas.
    }
    \label{fig:xray_spec}
\end{figure}

For Nucleus A, which has sufficient counts (76 and 82 net counts in two observations), we perform a spectral analysis, adopting C-statistics, which is suitable for the low-count regime \citep{cash79}. No significant spectral variation is found between the two observations, hence the two spectra are jointly fitted with all model parameters linked. We first try a single absorbed power-law model, i.e., $\rm{phabs \times (zphabs \times pow)}$ in XSPEC, where $\rm{phabs}$ represents the Galactic foreground absorption (fixed at 3$\times$10$^{20}$ cm$^{-2}$) and $\rm{zphabs}$ accounts for internal absorption to the AGN. This model, however, provides a poor fit to the spectra. Thus we introduce an additional power-law component to account for potential contribution from unresolved X-ray binaries and/or hot gas within the extraction aperture, i.e., $\rm{phabs \times (zphabs \times pow+pow)}$ in XSPEC. This phenomenological model provides an acceptable fit to the spectra, and the best-fit photon-index and absorption column density are typical of Type 2 Seyferts. Figure \ref{fig:xray_spec} shows the spectra and best-fit model for A. Table \ref{tab:xray} lists the X-ray measurements.

For Nuclei B and C, which have too few net counts for a spectral analysis, we estimate their X-ray spectral properties using hardness ratios (HR), defined as ${\rm HR} \equiv (H-S)/(H+S)$, where $H$ and $S$ are the number of net counts in the hard (2--8 keV) and soft (0.5--2 keV) X-ray bands. We use the Bayesian estimation of hardness ratios \citep{park06} to estimate HRs and uncertainties, appropriate for the low-count regime. Table \ref{tab:xray} lists the HR results and the estimated X-ray luminosity for B and C. We assume an absorbed power law with a photon index of $\Gamma_{{\rm X}}=1.7$ and an intrinsic absorption column density of $N_{\rm H}$=10$^{22}$ cm$^{-2}$ to calculate the luminosities for B and C based on their net count rates. The measured HR for B and C suggest softer spectra than the model assumption, although the inferred luminosity difference is insignificant. Strictly speaking, the estimated X-ray luminosities should be taken as lower limits considering the possibility of significantly underestimating the true absorbing column in the dusty merger due to the poor constraints from the low X-ray counts.

%%%%%%%%%%%%%%%%%%%%%%%%%%%%%%%%%%%%%%%%%%%%%%%%%%%%%%%%%%%%%%%%
%\newgeometry{margin=1cm}
%\begin{landscape}
\begin{table*}
\caption{X-ray measurements for \obj\ }
%\centering
\begin{tabular}{cccccccccc}
\hline
\hline
& & Flux$_{0.5-2\,{\rm keV}}$ & Flux$_{2-8\,{\rm keV}}$ & & & $N_{{\rm H}}$ & log $L_{0.5-2\,{\rm keV}}$ & log $L_{2-8\,{\rm keV}}$ & log $L_{2-10\,{\rm keV}}$ \\
ID & Counts & \multicolumn{2}{c}{(10$^{-6}$ photons cm$^{-2}$ s$^{-1}$)}  & HR & $\Gamma_{{\rm X}}$ & (10$^{22}$ cm$^{-2}$) & (erg s$^{-1}$) & (erg s$^{-1}$) & (erg s$^{-1}$)  \\
(1) & (2) & (3) & (4) & (5) & (6) & (7) & (8) & (9) & (10) \\
\hline
A & 178.8 & 6.76$^{+0.74}_{-0.73}$ & 6.35$^{+0.73}_{-0.72}$ & $-$0.05$^{+0.07}_{-0.07}$ & 1.12$^{+0.63}_{-0.48}$ & 0.66$^{+1.18}_{-0.66}$ & 41.67$^{+0.04}_{-0.04}$ & 42.01$^{+0.04}_{-0.05}$  &  42.13$^{+0.04}_{-0.05}$ \\
B & 17.7 & 1.17$^{+0.35}_{-0.30}$ & $<$0.77 & $-$0.80$^{+0.06}_{-0.16}$ & - & - & 41.27$^{+0.11}_{-0.13}$ & $<$40.90 & $<$40.99 \\
C & 8.3 & 0.49$^{+0.26}_{-0.21}$ & $<$0.76 & $-$0.73$^{+0.06}_{-0.27}$ & - & - & 40.90$^{+0.18}_{-0.23}$ & $<$40.90 & $<$40.98  \\
\hline
\end{tabular}
\tablecomments{
Column 1: same as Column 1 in Table \ref{tab:host}. 
Column 2: observed total net counts in 0.5--8 keV band. 
Columns 3 \& 4: observed photon flux. 
Column 5: hardness ratio HR $\equiv(H-S)/(H+S)$, where $H$ and $S$ are the number of counts in the hard and soft X-ray bands, respectively. 
Column 6: best-fit photon index assuming a power-law model where $n(E)\propto E^{-\Gamma_{\rm X}}$. 
Column 7: best-fit intrinsic column density assuming a power-law model. 
Columns 8--10: unabsorbed luminosity. For A, the luminosity is derived from the spectral fit, whereas for B and C it is inferred assuming an absorbed power law with a photon index of $\Gamma_{\rm X}=1.7$ and an intrinsic absorption column density of $N_{{\rm H}}=10^{22}$ cm$^{-2}$. All quoted errors are at 1 $\sigma$ confidence level, while the upper limits are at 3 $\sigma$.
}
\label{tab:xray}
\end{table*}
%\end{landscape}
%\restoregeometry
%%
%%

\subsection{VLA Radio 9.0 GHz Imaging, Radio Data Reduction, and Data Analysis}\label{subsec:vla}

\obj\ has an integrated 1.4 GHz flux density of 38.6$\pm$1.2 mJy from NVSS \citep{Condon1998} and 35.4$\pm$1.8 mJy\footnote{FIRST error bars need to be increased by 5\% in quadrature \citep{Perley2017}.} from FIRST \citep{White1992}. The FIRST source is centered on Nucleus A with a peak flux density of 30.2$\pm$1.5 mJy/beam, but the resolution is insufficient to separate emission from each individual nucleus.

To resolve the triple nucleus, we observed \obj\ with the VLA in its A configuration at 9.0 GHz on 2012 October 6 under project code SD0279 (PI Liu). The total exposure time was 1 hr with an on-source time of 0.5 hr. A total of 27 antennas were used. We reduced the new VLA data following standard procedures \citep{Wrobel2014a} using the Common Astronomy Software Aplications (CASA) package \citep{McMullin2007}. The beam size is 0\farcs19$\times$0\farcs19. Nuclei A and C were detected with peak flux densities of 2.58 mJy/beam and 0.348 mJy/beam and integrated flux densities of 5.52$\pm$0.17 mJy and 0.472$\pm$0.014 mJy, respectively.\footnote{VLA error bars need to be increased by 3\% in quadrature \citep{Perley2017}.} We have also detected some features with peak flux densities of 0.05--0.07 mJy/beam between Nucleus A and C, which could be galactic-scale star formation and/or AGN outflows. A 3$\sigma$ upper limit of 0.018 mJy/beam was set for Nucleus B.

We extrapolate our VLA 9.0 GHz measurements to 1.4 GHz assuming $f_{\nu}\propto\nu^{-0.5}$, which is canonical for Seyfert galaxies \citep{ho01}. Our approach is conservative because synchrotron self-absorption could produce a flatter spectral index. A flatter spectral index would result in a lower 1.4 GHz flux. This would in turn translate into a lower radio-based SFR estimate, which would make the case for an AGN even stronger (i.e., to explain the observed X-ray emission). For Nucleus B, we also explore an alternative model assuming $f_{\nu}\propto\nu^{-0.59}$ which is appropriate for star-forming galaxies \citep{Klein2018}, assuming the emission is dominated by nonthermal synchrotron. Again our approach is conservative because we assume the average spectral index for the nonthermal synchrotron emission observed in star-forming galaxies. Adopting the thermal free-free emission instead would result in a much flatter spectral index. Table \ref{tab:vla} lists our radio measurements.

%\newgeometry{margin=1cm}
\begin{table*}
\begin{center}
\caption{VLA radio measurements for \obj\  }
%\centering
\begin{tabular}{ccccccc}
  \hline
  \hline
 & S$^{{\rm Peak}}_{{\rm 9.0\,GHz}}$ & S$^{{\rm Int}}_{{\rm 9.0\,GHz}}$ & log L$_{{\rm 9.0\,GHz}}$ &  log L$_{{\rm 1.4\,GHz}}$ &  log L$^{{\rm SF}}_{{\rm 1.4\,GHz}}$ & SFR$_{{\rm 1.4\,GHz}}$ \\
ID & (mJy/beam) & (mJy) & (erg s$^{-1}$ Hz$^{-1}$) & (erg s$^{-1}$ Hz$^{-1}$) &  (erg s$^{-1}$ Hz$^{-1}$) & ($M_{\odot}$ yr$^{-1}$) \\
(1) & (2) & (3) & (4) & (5) & (6) & (7) \\
\hline
A & 2.58   & 5.52$\pm$0.17 & 30.09 & 30.49 & $\lesssim$29.49 & $\lesssim$20 \\
B & $<$0.018 & $<$0.018 & $<$27.61 & $<$28.08 & $<$28.08 & $<$0.8 \\
C & 0.348   & 0.472$\pm$0.014 & 29.03 & 29.43 & $\lesssim$28.43 & $\lesssim$2 \\
\hline
\end{tabular}
\end{center}
\tablecomments{
Column 1: same as Column 1 in Table \ref{tab:host}. 
Column 2: VLA 9.0 GHz peak flux density. 
Column 3: VLA 9.0 GHz integrated flux density in a 0\farcs5 radius aperture. VLA error bars have been increased by 3\% in quadrature \citep{Perley2017}. 
Column 4: rest-frame 9.0 GHz luminosity density. 
Column 5: rest-frame 1.4 GHz luminosity density extrapolated from L$_{{\rm 9.0\,GHz}}$ assuming $f_{\nu}\propto\nu^{-0.5}$, which is canonical for Seyfert galaxies \citep{ho01}. This is a conservative estimate because synchrotron self absorption could produce a flatter spectral index. For Nucleus B, we adopt an even more conservative model assuming $f_{\nu}\propto\nu^{-0.59}$, which is the average spectral index for nonthermal synchrotron observed in star-forming galaxies \citep{Klein2018}. Assuming a thermal free-free emission instead would produce a flatter spectral index. 
Column 6: estimated upper limit for the 1.4 GHz luminosity density contribution from star formation. For Nuclei A and C, we assume that $\lesssim$10\% of the total luminosity is from star formation (based on comparing the X-ray and {\it HST} $U$-band estimates). For Nucleus B, we adopt the most conservative estimate using the strict upper limit assuming that 100\% of the total emission is from star formation. 
Column 7: estimated SFR upper limit inferred from the 1.4 GHz luminosity density upper limit assuming the SFR$_{{\rm 1.4\,GHz}}$--L$_{{\rm 1.4\,GHz}}$ correlation \citep{Murphy2011}.
}
\label{tab:vla}
\end{table*}
%\restoregeometry
%\end{landscape}
%%
%%

%%%%%%%%%%%%%%%%%%%%%%%%%%%%%%%%%%%%%%%%%%%%%%%%%%%%%%%%%%%%%%%%%%%%%%%%%%%
\section{Results}\label{sec:result}

%%%%%%%%%%%%%%%%%%%%%%%%%%%%%%%%%%%%%%%%%%%%%%%%%%%%%%%%%%%%%%%%
%\newgeometry{margin=1cm}
%\begin{landscape}
\begin{table*}
\caption{Dust Extinction, Star Formation, and Intrinsic Luminosity Estimates }
%\centering
\begin{tabular}{cccccccccccc}
\hline
\hline
& $E(B-V)$ & $A_u$ & $A_z$ & log $L_{u,\,{\rm c}}$ & SFR$_{u,\,{\rm c}}$ & & SFR$_{{\rm d}}$ & log $L^{{\rm SF}}_{0.5-2\,{\rm keV}}$ & log $L^{{\rm SF}}_{2-10\,{\rm keV}}$ & log $L^{{\rm gal}}_{{\rm HX}}$ & log $L_{{\rm [O\,III],c}}$ \\
ID & (mag) & (mag) & (mag) & \bigg($\frac{{\rm erg}}{{\rm s\,Hz}}$\bigg) & \bigg($\frac{M_{\odot}}{{\rm yr}}$\bigg) & $D$(4000) & \bigg($\frac{M_{\odot}}{{\rm yr}}$\bigg) & (erg s$^{-1}$) & (erg s$^{-1}$) & (erg s$^{-1}$) & (erg s$^{-1}$)  \\
(1) & (2) & (3) & (4) & (5) & (6) & (7) & (8) & (9) & (10) & (11) & (12) \\
\hline
A$^{\dagger}$ & 0.51 & 2.50 & 0.74 & 29.0 & 8 & - & - & 40.57 & 40.61 & 40.49 & 43.10  \\
B$^{\dagger}$ & 0.82 & 4.01 & 1.19 & 29.8 & 70 & - & - & 41.48 & 41.53 & 41.06 & 42.72  \\
B$^{\ddagger}$ & 0.39 & 1.90 & 0.56 & 29.0 & 7 & 1.14 & 0.2 & [39.0, 40.5] & [39.0, 40.5] & [39.1, 40.1] & 41.66  \\
C$^{\dagger}$ & 0.56 & 2.75 & 0.81 & 28.5 & 2 & - & - & 39.88 & 39.95  & 39.79 &  42.39  \\
C$^{\ddagger}$ & 0.54 & 2.64 & 0.78 & 28.4 & 2 & 1.24 & 0.1 & [38.7, 40.0] & [38.7, 40.0] & [39.5, 39.8] & 41.63  \\
\hline
\end{tabular}
\tablecomments{
Column 1: same as Column 1 in Table \ref{tab:host}. The dagger indicates our own measurements from DIS slit spectra, and the double cross denotes those from the MPA-JHU SDSS DR7 value added catalog based on SDSS fiber spectra. 
Column 2: color excess inferred from \halpha /\hbeta\ using the Balmer decrement method, assuming the intrinsic case B values of 2.87 for $T = 10^4$ K \citep{osterbrock89} and the extinction curve of \citet{cardelli89} with $R_{V}=3.1$. 
Columns 3 and 4: SDSS rest-frame $u$- and $z$-band dust extinction estimated using the Balmer decrement method. 
Column 5: SDSS $u$-band luminosity density in rest frame (corrected for dust extinction) estimated using $k$-corrected $U$-band magnitude measured from \hst\ imaging. Column 6: SFR estimate inferred from $L_{u,{\rm c}}$ using Equation \ref{eq:sfr_lu}. 
Column 7: 4000-{\AA} break using the narrow definition \citep{Balogh1998}. 
Column 8: SFR estimate inferred from $D$(4000) assuming the empirical correlation observed in SDSS star-forming galaxies \citep{brinchmann04}. 
Columns 9 and 10: X-ray luminosities inferred from SFR$_{u,{\rm c}}$ using Equation \ref{eq:lx_sfr}. 
Column 11: galaxy-wide 2--10 keV luminosity (i.e., the combined emission from high-mass X-ray binaries and low-mass X-ray binaries) inferred from stellar mass and SFR assuming the empirical correlation \citep{Lehmer2010} observed in local luminous infrared galaxies given by Equation \ref{eq:lx_mass_sfr}. 
Column 12: \OIIIb\ luminosity corrected for dust extinction estimated using the Balmer decrement.
}
\label{tab:sf_dust}
\end{table*}
%\end{landscape}
%\restoregeometry
%%
%%

\subsection{Host-galaxy Internal Dust Extinction}\label{subsec:dust} 

We estimate the extinction correction using the Balmer decrement method \citep{osterbrock89} based on the emission-line intensity ratio \halpha /\hbeta\ measured from our spatially resolved APO/DIS spectra (Table \ref{tab:line}). We assume a dust-screen model with the intrinsic Case B value of \halpha /\hbeta =2.87 for $T=10^4$ K and the extinction curve of \citet{cardelli89} with $R_V$=3.1. To robustly measure \halpha /\hbeta , we have carefully subtracted the host-galaxy stellar continuum to minimize its error induced by Balmer absorption (Figure \ref{fig:spec}). 

Table \ref{tab:sf_dust} lists the estimated color excess and SDSS $u$- and $z$-band dust extinctions for each nucleus. Typical 1$\sigma$ statistical errors ($<$0.1 mag) are insignificant compared to systematic errors, which are difficult to quantify given the unknown dust geometry. Our baseline dust-screen model is simplistic. We neglect the effects of metallicity and ionization state in the approximation of assuming a fixed unattenuated Case B ratio. The metallicity-dependence of the Case B ratio would cause an overestimate of the dust attenuation by up to $\sim$0.5 mag at rest frame 6563 {\AA} for the most metal rich galaxies \citep{brinchmann04}.

\begin{figure}
  \centering
    \includegraphics[width=0.5\textwidth]{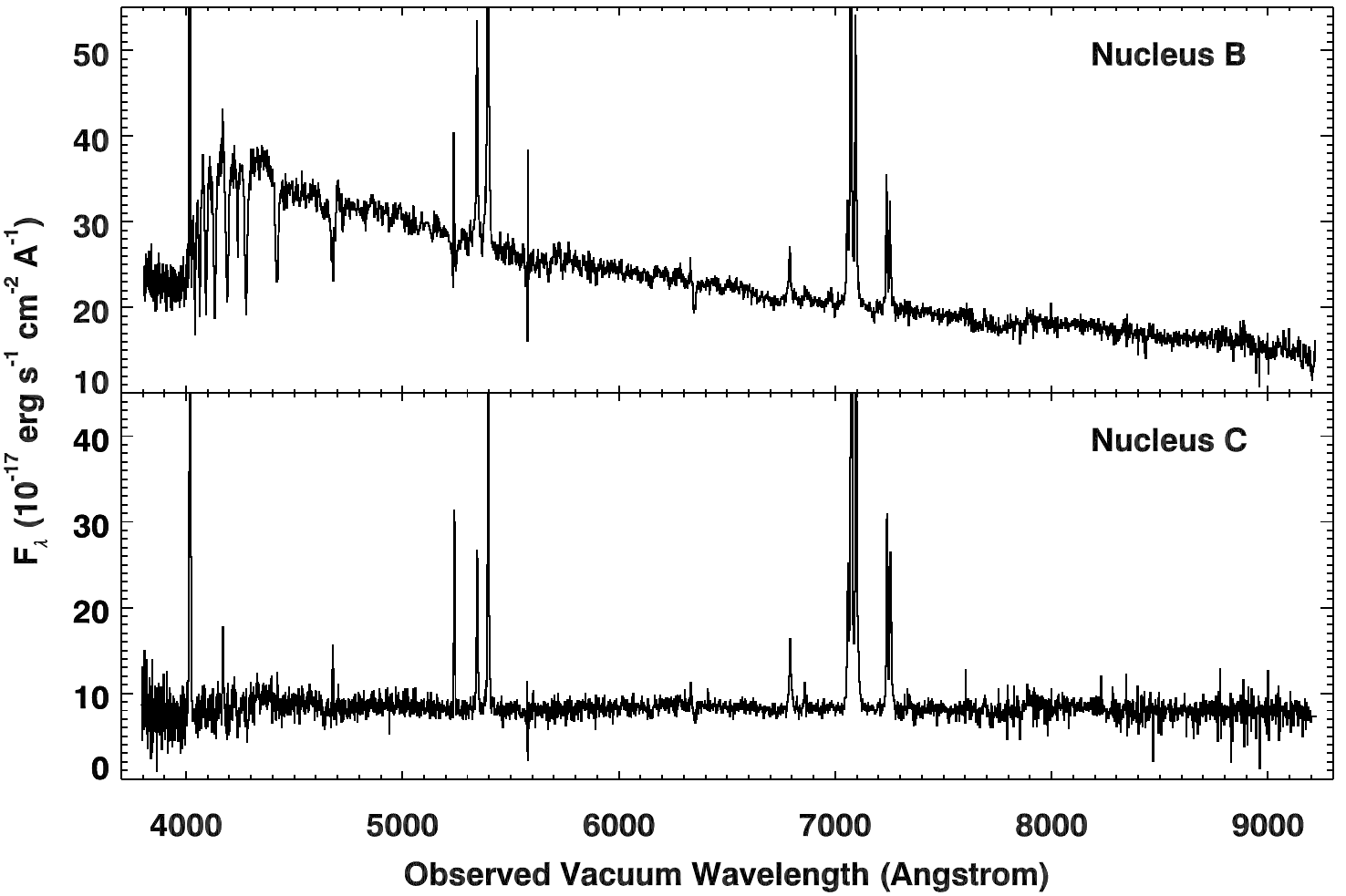}
    \caption{
    Archival SDSS spectra for Nuclei B and C. 
    The fiber aperture of these spectra is a 3$''$ diameter circle. Nucleus B shows strong Balmer absorption from post-starburst stellar populations. Unlike our APO 3.5m/DIS spectra, SDSS extends farther to the $u$ band that contains the Balmer absorption series. This series is helpful in breaking the degeneracy between strong Balmer absorption and Balmer emission for H$\alpha$ and H$\beta$ for Nucleus B. 
    }
    \label{fig:sdss_spec}
\end{figure}

Table \ref{tab:sf_dust} also lists the measurements based on the SDSS fiber spectra for Nuclei B and C. While the SDSS spectroscopic aperture is larger than that our APO/DIS spectra (and is therefore more contaminated by circumnucleus emission), the SDSS wavelength coverage is larger than that of APO/DIS spectra. In particular, the SDSS spectrum of Nucleus B shows strong Balmer absorption from a post-starburst stellar population (Figure \ref{fig:sdss_spec}). The strong Balmer absorption series in the blue end of the spectrum are crucial in helping to break the degeneracy between Balmer absorption and Balmer emission. On the other hand, our DIS spectrum does not cover the blue side, which contains the Balmer absorption series. Because of the limited coverage, it is likely that we have underestimated the absorption components (and overestimated the emission components) in H$\alpha$, H$\beta$, and H$\gamma$ in the DIS measurements. While the effects on the BPT ratios are likely to be minor, this would bias the Balmer decrement (Table \ref{tab:line}), and by extension, the dust extinction (Table \ref{tab:sf_dust}) estimate high, resulting in an overestimated SFR estimate. In view of this caveat, we adopt the SDSS-based measurements as our fiducial estimates for Nucleus B.

\subsection{Host-galaxy Stellar Mass}\label{subsec:stellarmass} 

We estimate the total stellar mass for the host of each nucleus using the $k$-corrected SDSS $z$-band absolute magnitude at rest frame after correction for dust extinction, $M_{z,\,{\rm c}}$, and the stellar mass-to-light ratio $M/L$ inferred from SDSS $u-z$ color ($k$-corrected to the rest frame). We adopt the empirical relation given by \citet{Bell2003}
\begin{equation}\label{eq:stellar_mass}
\mathrm{log_{10}}\left(\frac{M_*}{M_\odot}\right) = -0.4 \, (M_z - M_{z, \odot})
- 0.179 + 0.151 \, (u-z),
\end{equation}
where $M_*$ is the galaxy stellar mass in solar units and $M_{z, \odot}=4.51$ is the absolute magnitude of the Sun in the $z$ band \citep{Blanton2007}. Typical systematic uncertainties from galaxy age, dust, and bursts of star formation are $\sim$0.1 dex for optical $M/L$ and are larger for bluer galaxies \citep{Bell2003}. Uncertainties in stellar evolution are likely to dominate the systematic error in stellar mass, which is expected to be $\sim$0.3 dex \citep{Conroy2013}.

We also estimate the stellar mass contained within the same region as the X-ray extraction aperture for each nucleus, $M_{\ast,\,{\rm ap}}$. Along with the specific SFR inferred from optical continuum spectral index, $M_{\ast,\,{\rm ap}}$ is used to obtain an independent estimate for the SFR (see below for details). We estimate $M_{\ast,\,{\rm ap}}$ using a similar approach for estimating the total stellar mass, except that we use the aperture magnitudes $m_U$ and $m_Y$ and the aperture $u-z$ color estimated by applying $k$-corrections to $m_U$ and $m_Y$. Assuming a flat local spectrum between $u$ and $U$ and between $z$ and $Y$, the $k$-corrections are given by $m_z$ = $m_Y$ + 0.36 and $m_u$=$m_U$ $-$ 0.13. Table \ref{tab:host} lists our results.

%%%%%%%%%%%%%%%%%%%%%%%%%%%%%%%%%%%%%%%%%%%%%%%%%%
\subsection{Host-galaxy Nuclear SFR}\label{subsec:sfr} 

We estimate the host-galaxy nuclear SFR from the \hst -measured $U$-band luminosity following the method of \citet{Liu2013}. We choose the \hst\ $U$-band over the SDSS $u$-band imaging because the former resolves the emission from each nucleus better. We measure the \hst\ $U$-band magnitude for each nucleus using aperture photometry. We adopt the same aperture as in the X-ray source extraction to minimize any aperture effects. We then convert the \hst\ $U$-band magnitude to rest-frame SDSS $u$-band using a $k$-correction estimated by comparing the SDSS $u$-band aperture magnitude with the \hst\ $U$-band magnitude from the same aperture as the SDSS.  Table \ref{tab:sf_dust} lists $L_{u,\,{\rm c}}$, the intrinsic $u$-band luminosity density corrected for internal dust extinction, and the inferred SFR.   

The $U$ band covers the rest frame 2870--3380 \AA\ for \obj . For Type 2 Seyferts where the AGN power law is obscured in the UV/optical, the luminosity in this range is dominated by continuum photosphere emission of young stellar populations in the host galaxy, providing a useful indicator for SFR \citep{cram98}. It is relatively free from strong emission lines from ionized gas, AGN light from the obscured nucleus scattered into our line of sight by dust and/or gas \citep{zakamska06}, considering the moderate AGN luminosity and the absence of a broad H$\beta$ or a broad H$\alpha$ in the optical spectrum \citep{Liu2009}, and the nebular continuum from ionized gas in the AGN emission-line region \citep{osterbrock89}. Accounting for possible contamination from all of these non-star formation related processes would lower the $U$-band--based SFR estimates and strengthen the case of AGN excitation.   

To infer the SFR from $L_{u,\,{\rm c}}$, we have adopted the empirical calibration \citep{hopkins03} obtained based on 2625 SDSS star-forming galaxies, given by
\begin{equation}\label{eq:sfr_lu}
\frac{{\rm SFR}_{u,\,{\rm c}}}{M_{\odot}~{\rm yr}^{-1}} = \bigg(\frac{L_{u,\,{\rm c}}}{1.81 \times 10^{28}~
{\rm ergs~s}^{-1} {\rm Hz}^{-1}}\bigg)^{1.186},
\end{equation}
which is valid for 2$\times$ 10$^{28}$ ergs s$^{-1}$ Hz$^{-1}$ ${\lesssim}L_{u,\,{\rm c}} {\lesssim}$10$^{30}$ ergs s$^{-1}$ Hz$^{-1}$. Equation \ref{eq:sfr_lu} has been calibrated against \halpha -based SFR with an rms scatter of 0.13 dex. The calibrated $u$-band SFR has been shown to be consistent with the radio 1.4 GHz based SFR with an rms scatter of 0.23 dex \citep{hopkins03}.

We compare the $U$-band-based SFR with an independent estimate inferred from the 4000 {\AA} break $D$(4000) using the narrow definition \citep{Balogh1998}. SFR indicators based on optical emission lines such as H$\alpha$ luminosity cannot be easily applied to \obj\ given the likely significant contamination from AGN excitation. We adopt $D$(4000) because all three nuclei in \obj\ are Type 2, i.e., obscured Seyferts, in which the AGN power law is obscured in the optical and the observed continuum is dominated by host-galaxy starlight. From $D$(4000), we infer the specific SFR, i.e., SFR/$M_{\ast}$, assuming the empirical calibration based on SDSS star-forming galaxies \citep{brinchmann04}. We then estimate the SFR by multiplying the specific SFR with the stellar mass within the same region as the X-ray extraction aperture, $M_{\ast,\,{\rm ap}}$ (Table \ref{tab:host}). The $D$(4000) based SFR, SFR$_{{\rm d}}$, is consistent with the $U$-band-based SFR within uncertainties (Table \ref{tab:sf_dust}), lending support to our SFR estimates.

For Nucleus B, we adopt the SDSS-based SFR estimates as our fiducial value. The DIS-based SFR is likely to be significantly overestimated due to the likely overestimated dust extinction correction. The more moderate SFR estimate is also more in line with the total (i.e., from three nuclei combined) SFR (${\sim}$20 M$_\odot$ yr$^{-1}$) inferred from the total IR luminosity \citep[log($L_{{\rm IR}}/L_{\odot}$=11.43$\pm$0.03;][]{Pfeifle2019} as well as the strict upper limit from its radio luminosity density (see \S \ref{subsec:vla} for details), both of which are much less subject to uncertainties caused by the dust-attenuation correction.

%%%%%%%%%%%%%%%%%%%%%%%%%%%%%%%%%%%%%%%%%%%%%%%%%%
\subsection{X-Ray Contribution from Nuclear Star Formation}\label{subsec:sf_xray} 

The estimated unabsorbed hard X-ray luminosities for the three nuclei are close to or below 10$^{42}$ erg s$^{-1}$, the characteristic upper limit for X-ray luminous starburst galaxies \citep{zezas01}. To determine whether the X-ray luminosity is dominated by star-formation-induced processes or by AGN excitation, X-ray spectral properties provide a discriminating diagnostic, although uncertainties in the X-ray measurements are too large to draw a firm conclusion. 

Alternatively, we derive independent estimates on the expected X-ray luminosity due to star-formation-related processes in each nucleus and compare with the total X-ray luminosity to assess whether AGN is needed as an additional excitation source. We estimate the X-ray contribution from nuclear star formation using SFR$_{u,\,{\rm c}}$, the $U$-band-based aperture SFR corrected for dust extinction. We adopt the empirical relation given by
\begin{equation}\label{eq:lx_sfr}
    \begin{split}
L^{{\rm SF}}_{0.5-2\, {\rm keV}} = 4.5 \times 10^{39} \frac{{\rm SFR}}{M_{\odot}~{\rm
yr}^{-1}} {\rm ergs~s}^{-1}, \\
L^{{\rm SF}}_{2-10\, {\rm keV}} = 5.0 \times 10^{39} \frac{{\rm SFR}}{M_{\odot}~{\rm
yr}^{-1}} {\rm ergs~s}^{-1},
\end{split}
\end{equation}
which is calibrated using 23 nearby star-forming galaxies \citep{grimm03,ranalli03}. The rms scatter is 0.27 dex in the soft and 0.29 dex in the hard X-ray band. Table \ref{tab:sf_dust} lists the resulting estimates.

Our estimates of the expected X-ray luminosity due to star-formation-related processes in the nucleus may be highly uncertain due to systematics in the model assumptions. The uncertainty is likely dominated by systematic errors in the $U$-band extinction correction. The Balmer decrement is estimated using emission-line measurements for each nucleus from our APO/DIS spectra. Given the lower angular resolution and projection effects, the aperture of the emission-line measurement does not exactly match with the \hst\ $U$-band measurement, causing errors due to aperture mismatch. The reddening uncertainty due to the aperture mismatch is likely to be $<$1 mag \citep{Liu2013}. More importantly, however, the color excesses we derived using the Balmer decrement method may not represent the true dust attenuation. For example, dust could concentrate on scales smaller than where the Balmer lines are emitted. Patchy and optically thick dust clouds could result in high extinction with little reddening. 

Our inferred star-formation-related X-ray luminosity is not sensitive to the initial mass function (IMF) uncertainty, assuming no large systematic IMF variations among star-forming galaxies \citep{scalo86,kroupa01,chabrier03}. A Salpeter IMF \citep{salpeter55} is assumed with mass limits of 0.1 $M_{\odot}$ and 100 $M_{\odot}$ in the calibration of SFR from $U$-band luminosity (Equation \ref{eq:sfr_lu}). The adopted SFR--X-ray-luminosity relation \citep[Equation \ref{eq:lx_sfr};][]{ranalli03} is calibrated under the same assumptions about the IMF and mass range \citep{kennicutt98}.

We double-checked our results using the empirical correlation between galaxy-wide 2--10 keV luminosity L$^{{\rm gal}}_{{\rm HX}}$ (assumed to be the combined emission from high-mass X-ray binaries and low-mass X-ray binaries), SFR, and stellar mass from \citet{Lehmer2010}. This calibration is based on a sample of 17 local luminous IR galaxies. The relation is given by
\begin{equation}\label{eq:lx_mass_sfr}
    L^{{\rm gal}}_{{\rm HX}} = \alpha M_{\ast} + \beta {\rm SFR},
\end{equation}
where $\alpha=(9.05\pm0.37)\times10^{28}$ erg s$^{-1}$ $M_{\odot}^{-1}$ and $\beta=(1.62\pm0.22)\times10^{39}$ erg s$^{-1}$ ($M_{\odot}$ yr$^{-1}$)$^{-1}$. The stellar mass and SFR-derived hard X-ray luminosities are consistent with the SFR-derived hard X-ray estimates within the uncertainties (Table \ref{tab:sf_dust}), lending confidence to our estimates.

%%%%%%%%%%%%%%%%%%%%%%%%%%%%%%%%%%%%%%%%%%%%%%%%%%
\subsection{Photoionization Estimates}\label{subsec:photo} 

We use \OIIIc /\NIIc\ and \OIIIc /\SIIab\ (corrected for reddening) to quantify the level of ionization in the emission-line gas. Both line flux ratios increase as the level of ionization increases. The dimensionless ionization parameter, $U$, defined as the ratio of the density of ionizing radiation at some position $i$ to the electron density $n_e$, is given by
\begin{equation}
    U_{i} \equiv \frac{Q}{4\pi d^2_i n_e c}, 
\end{equation}
where $Q$ is the number of ionizing photons per second emitted by the ionizing source, $d_i$ is the distance to position $i$ from the ionizing source, and $c$ is the speed of light. In the case of a central AGN, $U_{i}$ decreases as $d_{i}$ increases unless $n_e$ also decreases at a faster rate than $1/d^2$. The \OIIIc /\NIIc\ and \OIIIc /\SIIab\ line ratios depend on four parameters: ionization parameter, spectral index (which characterizes the hardness of the ionizing spectrum), gas density, and metallicity. The line ratios will increase outward with an increase in density, a decrease in metallicity, or an increase in ionization parameter. 

Since Nuclei B and C have both SDSS and APO/DIS spectra, we can compare their separately measured emission-line ratios to study the ionization gradient. Systematic differences in the absolute flux calibration between SDSS and APO/DIS do not affect line-ratio comparisons. The SDSS spectroscopic aperture is a 3$''$ diameter fiber, whereas the APO/DIS aperture is roughly 1\farcs5$\times$2$''$, i.e., probing a more central region than the SDSS does. As listed in Table \ref{tab:line}, the APO 3.5m/DIS measurements of \OIIIc /\NIIc\ and \OIIIc /\SIIab\ are both larger than those from SDSS. This suggests that the level of ionization decreases with increasing distance to the nucleus for both B and C. The line-ratio gradients cannot be explained by either an outward-increasing density (which is unphysical; the observed electron densities from the SDSS and APO/DIS measurements are consistent within the uncertainties, further ruling this possibility out), or an outward-decreasing metallicity (because the required metallicity would be too high; the line ratios also become insensitive to metallicity at low values, which are expected for both B and C given their low stellar masses). This outward-decreasing ionization parameter is consistent with AGN, in contrast to the outward-increasing ionization parameter expected for distributed ionizing sources such as post-AGB stars, which can produce LINER-like emission-line ratios \citep[e.g.,][]{Yan2012}.

We also compare the line ratios \OIII /\NII\ and \OIII /\SII\ of B against those of A and C (all based on the APO/DIS measurements, which characterize the more central region) to assess whether the gas associated with Nucleus B could be solely ionized by the AGN in Nucleus A and/or C instead of being primarily ionized by its own AGN. The \OIII /\NII\ and \OIII /\SII\ line ratios of Nucleus B are both comparable to those of Nucleus A and are higher than those of Nucleus C. Based on arguments similar to the above, this comparison suggests that an additional central ionizing source in B is needed and it is not solely ionized by A and/or C. Our conclusion is robust against uncertainties in the dust-reddening correction of the line ratios \OIII /\NII\ and \OIII /\SII .

In summary, the line ratios \OIII /\NII\ and \OIII /\SII\ are most sensitive to the ionization parameter that characterizes the level of ionization. The comparison between the SDSS and APO/DIS-based measurements suggests an outward-decreasing ionization parameter in both B and C, in favor of the AGN hypothesis. The high \OIII /\NII\ and \OIII /\SII\ line ratios of B as compared against A and C is also in favor of an additional AGN in B, rather than being solely ionized by A and/or C.  

\subsection{Nature of the Nuclear Ionizing Sources}\label{subsec:nature}

First, \chandra\ ACIS-S detects all three nuclei in the soft X-rays (0.5--2 keV) as compact sources whose spatial profiles are consistent with the PSF (Figure \ref{fig:xray_radio}). The soft X-ray center is consistent with that of the \hst\ $Y$-band nucleus for all three nuclei. Nuclei A and C are both detected as compact (beam size 0\farcs19$\times$0\farcs19) radio point sources by the VLA in 9.0 GHz (Figure \ref{fig:xray_radio}). The radio center is consistent with that of the \hst\ $Y$-band nucleus for both nuclei. 

\chandra\ also detects Nucleus A as a compact point source in the hard X-rays (2--8 keV) whose center is consistent with the \hst\ $Y$ band nucleus. Its estimated unabsorbed hard X-ray (2--10 keV) luminosity is 10$^{42.13^{+0.04}_{-0.05}}$ erg s$^{-1}$, which is comparable to the most X-ray luminous starburst galaxies \citep{zezas01}. However, the hard X-ray luminosity cannot be explained by pure star-formation-related processes, because the SFR in the host of A is only modest, estimated as $\sim$8 $M_{\odot}$ yr$^{-1}$ (corrected for dust attenuation). The SFR is inferred from \hst\ $U$-band luminosity (corrected for external dust extinction) assuming the empirical calibration based on star-forming galaxies \citep{hopkins03} and verified by an independent estimate based on the 4000 {\AA} break \citep{brinchmann04}, which is consistent with the independent constraint ($\lesssim$20 $M_{\odot}$ yr$^{-1}$) based on the VLA radio 9.0 GHz luminosity. On the other hand, the hard X-ray luminosity may be $\sim$10 times higher than our baseline estimate considering systematic uncertainties in the model assumption, making the AGN case even stronger. The high hard X-ray luminosity, particularly when compared to its moderate SFR, unambiguously confirms A as an AGN.  

Second, while Nucleus C is undetected in the hard X-rays, it is detected as a compact (beam size 0\farcs19$\times$0\farcs19) radio source by the VLA in 9.0 GHz (Figure \ref{fig:xray_radio}). In addition, its SFR is estimated as $\sim$0.1--2 $M_{\odot}$ yr$^{-1}$ (corrected for dust attenuation) based on HST $U$-band luminosity and corroborated by the host-galaxy continuum spectral index, which is also consistent with the independent constraint ($\lesssim$2 $M_{\odot}$ yr$^{-1}$) based on the VLA radio 9.0 GHz luminosity assuming that $\lesssim$10\% of the radio luminosity comes from a nuclear starburst. This is too low to explain its soft X-ray luminosity by pure star-formation-related processes. Based on the empirical correlation between SFR and L$_{\rm X}$ \citep{ranalli03} observed in star-forming galaxies, the expected soft X-ray luminosity from pure star-formation-related processes is estimated to be L$^{{\rm SF}}_{0.5-2\,{\rm keV}}{\sim}$10$^{38.7}$--10$^{40.0}$ erg s$^{-1}$, which is about an order of magnitude lower than the total soft X-ray luminosity L$_{0.5-2\,{\rm keV}}{\sim}$10$^{40.90^{+0.18}_{-0.23}}$ erg s$^{-1}$. Its extinction-corrected $U$-band SFR is more than an order of magnitude lower than that expected from its radio luminosity, assuming it were 100\% star formation  (i.e., composed of nonthermal synchrotron emission and thermal bremsstrahlung (free-free) emission), according to the empirical correlation between SFR and $L_R$ \citep{bell03} observed in star-forming galaxies.

Finally, while Nucleus B is neither detected in the hard X-rays nor in the radio, its soft X-ray luminosity ($L_{0.5-2\,{\rm keV}}{\sim}$10$^{41.27^{+0.11}_{-0.13}}$ erg s$^{-1}$) exceeds the luminosity that we would expect from pure star-formation-related processes ($L^{{\rm SF}}_{0.5-2\,{\rm keV}}{\sim}$10$^{39.0}$--10$^{40.5}$ erg s$^{-1}$), which is derived from its HST $U$-band and continuum-index inferred SFR estimates ($\sim$0.2--7 $M_{\odot}$ yr$^{-1}$ corrected for dust attenuation). Furthermore, the new VLA image allows us to place a stringent radio luminosity upper limit, which independently sets a 3$\sigma$ SFR upper limit of SFR$<$0.8 $M_{\odot}$ yr$^{-1}$. This translates into an even tighter 3$\sigma$ upper limit for $L^{{\rm SF}}_{0.5-2\,{\rm keV}}{<}$10$^{39.6}$ erg s$^{-1}$, which is more than an order of magnitude lower than the observed total soft X-ray luminosity. The radio-based SFR estimate (and by extension, the soft X-ray luminosity caused by star-formation-related processes) is more robust against uncertainties in the dust-attenuation correction than the $U$-band or host stellar continuum-index estimates. This suggests contribution from an additional excitation source such as an AGN and/or shock heating. In addition, photoionization arguments suggest that its Seyfert-type BPT diagnostic line ratio is unlikely to be caused by pure star-formation-related process and/or shock heating, or primarily by the AGN in Nuclei A and/or C. The line ratios \OIII /\NII\ and \OIII /\SII\ are most sensitive to the ionization parameter that characterizes the level of ionization. Comparison between the SDSS (from a 3$''$ diameter aperture) and APO/DIS-based (roughly from a 1\farcs5$\times$2$''$ aperture, i.e., probing a more central region than the SDSS) line ratios suggests an outward-decreasing ionization parameter in both B and C, in favor of the AGN hypothesis. The high \OIII /\NII\ and \OIII /\SII\ line ratios of B as compared against A and C are also in favor of an additional AGN in B, rather than being solely ionized by A and/or C. Finally, while shocks could also produce excess of soft X-ray emission, the Seyfert-type optical emission-line ratio is difficult to accommodate in a purely shock-heating scenario.

In summary, our complementary multiwavelength follow-up observations lend further support to the triple BH scenario suggested by the optical slit spectroscopy. Taken together, our comprehensive observations strongly suggest that all three nuclei are AGN, making \obj\ the most unambiguous case known to host a kpc-scale trio of MBHs.

\section{Discussion}\label{sec:discuss}

%%%%%%%%%%%%%%%%%%%%%%%%%%%%%%%%%%%%%%%%%%%%%%%%%%
\subsection{Black Hole Mass} 

We estimate the BH mass for each nucleus inferred from the host-galaxy total stellar mass and bulge stellar velocity dispersion assuming the $M_{\bullet}$--$M_{\ast}$ and $M_{\bullet}$--$\sigma_{\ast}$ relations \citep{KormendyHo2013}. For the $M_{\bullet}$--$M_{\ast}$ relation, we prefer the host-galaxy total stellar mass over the bulge stellar mass because (i) there is empirical evidence that the correlation between BH mass and total stellar mass is tighter than that between BH mass and bulge stellar mass in low-redshift active galaxies \citep{KormendyHo2013}, and (ii) our bulge and disk decomposition may be uncertain, making the total stellar mass estimate more robust than the bulge stellar mass estimate. We adopt the relation given by
\begin{equation}\label{eq:bh_totalmass}
    \mathrm{log_{10}}\bigg(\frac{M_{\bullet}}{M_{\odot}}\bigg)=(7.45\pm0.08) + (1.05\pm0.11)\,\mathrm{log_{10}}\bigg(\frac{M_{\ast}}{10^{11}M_{\odot}}\bigg),
\end{equation}
which is calibrated based on a sample of 262 local broad-line AGN with virial BH mass estimates \citep{Reines2015}. The rms scatter in the relation is 0.55 dex, which includes the measurement error in the virial BH mass (0.50 dex) and the best-fit intrinsic scatter (0.24 dex). The stellar masses in the calibration sample of Equation \ref{eq:bh_totalmass} are based on the M/L ratios in \citet{Zibetti2009}, which assumes a Chabrier IMF. Our stellar mass uses the M/L ratios in \citet{Bell2003} assuming the scaled Salpeter IMF. The \citet{Bell2003} masses are systematically higher than the \citet{Zibetti2009} masses by 0.21 dex \citep{Reines2015}. To correct for this systematic offset, we subtract 0.21 dex from our stellar mass estimate as the input for Equation \ref{eq:bh_totalmass}. Table \ref{tab:host} lists the estimated BH mass for each nucleus from the $M_{\bullet}$--$M_{\ast}$ relation. 

For the $M_{\bullet}$--$\sigma_{\ast}$ relation, we adopt the relation given by
\begin{equation}\label{eq:bh_vdisp}
    \mathrm{log_{10}}\bigg(\frac{M_{\bullet}}{M_{\odot}}\bigg)=(7.96\pm0.03) + 4.02\,\mathrm{log_{10}}\bigg(\frac{\sigma_{\ast}}{200\,{\rm km}\,{\rm s}^{-1}}\bigg),
\end{equation}
which is calibrated based on a sample of 71 local active galaxies with virial BH mass estimates \citep{Greene2006}. The intrinsic scatter in the relation is $<$0.4 dex. Table \ref{tab:host} lists the estimated BH mass for each nucleus from the $M_{\bullet}$--$\sigma_{\ast}$ relation, which is consistent with the estimate from the $M_{\bullet}$--$M_{\ast}$ within the uncertainties, lending support to our BH mass estimates. While still broadly consistent with the stellar-mass-based estimates, the stellar velocity-dispersion-based estimates may result in overestimated BH masses, considering that stellar velocity dispersions are systematically higher in mergers than in isolated AGN host galaxies \citep[e.g.,][]{Liu2012}. Our estimates are also broadly consistent with independent constraints assuming the fundamental place relation of BH accretion \citep{Merloni2003,Gultekin2019}.

%%%%%%%%%%%%%%%%%%%%%%%%%%%%%%%%%%%%%%%%%%%%%%%%%%
\subsection{Merging Timescales}\label{subsec:timescale} 

We use Chandrasekhar dynamical friction timescale arguments to estimate characteristic timescales for the galaxy-galaxy mergers and the inspiraling timescales of the BHs. Assuming singular isothermal spheres for the host-galaxy density profiles and circular orbits, we estimate the dynamical friction timescale for a satellite galaxy with velocity dispersion $\sigma_{{\rm S}}$ inspiraling from radius $r$ in a host galaxy with velocity dispersion $\sigma_{{\rm H}}$ as \citep{Chandrasekhar1943}
\begin{equation}\label{eq:t_galfric}
t^{{\rm gal}}_{{\rm fric}} = \frac{2.7}{\ln \Lambda}\frac{r}{30\,{\rm kpc}}\bigg(\frac{\sigma_{{\rm H}}}{200\, {\rm km\,s^{-1}}}\bigg)^2\bigg(\frac{100\, {\rm km\,s^{-1}}}{\sigma_{{\rm S}}}\bigg)^3 ~{\rm Gyr},
\end{equation}
where $\Lambda$ is the Coulomb logarithm with typical values of 3${\lesssim}\ln \Lambda{\lesssim}$30 \citep{binney87}. $\ln \Lambda{\sim}$2 for equal-mass mergers \citep{dubinski99}. We assume $\ln\Lambda{\sim}$6. With A as the host and B and C as satellites, B and C will merge with A in $t_{{\rm fric}}{\sim}$56$^{+36}_{-28}$ Myr and $\sim$49$^{+77}_{-25}$ Myr, corrected for projection effects assuming random orientation. 

After merging together with their individual hosts, the three BHs will in-spiral to the center of the merged galaxy under dynamical friction with the stellar background. Assuming a singular isothermal sphere for the density distribution of the merged host galaxy, the dynamical friction timescale of a BH of mass $M_{\bullet}$ on a circular orbit of radius $r$ is estimated as
\begin{equation}\label{eq:t_bhfric}
t^{{\rm BH}}_{{\rm fric}} = \frac{19}{\ln \Lambda}\bigg(\frac{r}{5\,{\rm kpc}}\bigg)^2\frac{\sigma_{{\rm H}}}{200\,{\rm km\,s^{-1}}}\frac{10^8\,M_{\odot}}{M_{\bullet}}~{\rm Gyr}, 
\end{equation}
where $\ln \Lambda{\sim}$6 for typical values \citep{binney87}. We estimate $t_{{\rm fric}}{\sim}$1.9$^{+2.0}_{-0.9}$ Gyr and 1.5$^{+0.9}_{-1.0}$ Gyr for the BH in B and C to reach the center of the merger products to form a gravitationally bound triple BH system with A, assuming a radius of $r$=1 kpc at the start of the inspiraling. 

We have neglected the effects of friction by gas, which will accelerate the merger processes. On the other hand, we have neglected the effects of tidal stripping of stars in the satellites, which will delay the merger. We have also neglected the effect of BH accretion, which may change the masses of the BHs and the resulting merger timescales. The actual merger involving three stellar and BH components with gas friction and gas accretion onto three BHs is likely much more complicated. Numerical simulations with model parameters tailored to those of \obj\ are needed to make more realistic predictions of the subsequent merger of the triple system \citep{renaud10}. Regardless of the actual sequence of the mergers, the BHs are likely to form a gravitationally interacting triple system if the orbit of the third MBH decays rapidly enough before the first two BHs coalesce.

\section{Summary and Future Work}\label{sec:sum}

We have reported the discovery of \obj , the first case of a triple Type 2 Seyfert nucleus. It represents three active BHs hosted in a three-way galaxy merger. The target was selected from the largest sample of optically selected candidate AGN pairs based on the SDSS data. However, the archival SDSS data only have spectra for two of the three nuclei; the SDSS fibers were also too large to clearly separate the emission from the nucleus. To unambiguously determine the nature of the triple nucleus, we have conducted a comprehensive multiwavelength follow-up campaign. We summarize our main findings as follows:

\begin{itemize}

\item First, by conducting new spatially resolved optical spectroscopy, we have classified all three nuclei as Type 2 Seyferts based on the classical BPT diagram. This establishes J0849+1114 as the first known case of a triple Type 2 Seyfert nucleus.

\item Furthermore, to confirm the excitation mechanism of the triple Type 2 Seyfert nucleus, we have further conducted a complementary follow-up campaign including {\it Chandra} X-ray imaging spectroscopy, HST $U$- and $Y$-band imaging, and VLA radio imaging. These new comprehensive multiwavelength observations strongly suggest that J0849+1114 hosts a kpc-scale triple AGN. This represents the most robust evidence for three active MBHs in the process of merging. The discovery provides a unique verification of the hierarchical MBH-formation paradigm with important implications for the detection of low-frequency gravitational waves.

\item By modeling the host-galaxy photometry and internal dust extinction of \obj , we have estimated the stellar masses of the three merging components to be $\sim$10$^{11.3}M_{\odot}$, 10$^{10.0}M_{\odot}$, and 10$^{10.5}M_{\odot}$ for A, B, and C. Assuming the empirical correlation between BH mass and host total stellar mass observed in local broad-line AGN \citep{Reines2015}, the inferred BH masses are $\sim$10$^{7.5}M_{\odot}$, 10$^{6.4}M_{\odot}$, and 10$^{6.7}M_{\odot}$, consistent with independent estimates based on the host-galaxy stellar velocity dispersion within the uncertainties.

\end{itemize}

\obj\ provides a unique verification of the hierarchical merger-tree model for MBH formation \citep[e.g.,][]{volonteri03}. Smaller galaxies merge to form larger ones \citep{toomre72}. After galaxies merge, massive binary BHs should form through dynamical friction \citep{begelman80,milosavljevic01,yu02}. When a third galaxy enters before the first two BHs coalesce, an MBH triple should form \citep{valtonen96}. Triple MBHs are important for understanding galaxy formation. They could scour out stellar cores much larger than those formed around binaries \citep{hoffman07}. In addition, triple MBHs represent a unique laboratory for studying the chaotic dynamics of general relativity three-body interactions \citep{blaes02,merritt06}. Their end products may include a merger of all three BHs, formation of a highly eccentric binary, or even ejection of three free BHs \citep{lousto08}. The inner binary may have a very high eccentricity, releasing intense bursts of gravitational waves sought by pulsar-timing arrays and the Laser Interferometer Space Antenna \citep{amaro10}. Three-body interactions also represent one of the main mechanisms to solve the final-parsec problem for rapid binary coalescence \citep{valtonen96,blaes02,hoffman07}.

Despite significant merit and intense interest, direct evidence for triple MBHs is still lacking. Their separations (less than a few parsecs) are too small to resolve beyond the local universe with current facilities. Kpc-scale triple AGN represent a promising precursor of compact triple MBHs. Using a simple stellar dynamical friction argument (\S \ref{subsec:timescale}), we have estimated that the trio in \obj\ may form a bound MBH triple in $\lesssim$2 Gyr. 

Future much deeper X-ray observations with {\it Chandra} may detect and resolve the two fainter nuclei in the hard X-rays and characterize their X-ray spectral properties. Deeper VLA observations, now under way, may detect the faintest Nucleus B in the radio and measure its radio spectral index. {\it SOFIA}/FORCAST imaging may measure the mid-IR spectral energy distributions and marginally resolve the three nuclei. Finally, high spatial resolution, integral field unit spectroscopy in the optical and/or near-IR may map out the ionization and velocity fields around the triple AGN.

The selection of \obj\ is limited because its parent galaxy sample is based on the SDSS. Future large near-IR spectroscopic galaxy surveys \citep{Takada2014,TheMSEScienceTeam2019} may uncover a larger population of galaxies harboring triple AGN in the higher redshift universe where mergers are believed to be more frequent.

%------------------------------------------------------------------------------
\section*{Acknowledgements}

We thank S. Gilmore, M. Claussen, and N. Ferraro for help with our VLA observations, and the anonymous referee for a prompt and helpful report. X.L. acknowledges support by NASA through {\it Chandra} Award Number GO3-14104X issued by the {\it Chandra} X-ray Observatory Center, which is operated by the Smithsonian Astrophysical Observatory for and on behalf of NASA under contract NAS 8-03060. Support for Program number HST-GO-13112 was provided by NASA through a grant from the Space Telescope Science Institute, which is operated by the Association of Universities for Research in Astronomy, Incorporated, under NASA contract NAS5-26555. M.H. and Z.L. acknowledge support by the National Key Research and Development Program of China (2017YFA0402703). M.K. acknowledges support by NSFC Youth Foundation (No. 11303008) and by the Astronomical Union Foundation under grant No. U1831126. Y.S. acknowledges partial support from an Alfred P. Sloan Research Fellowship and NSF grant AST-1715579. This work makes extensive use of SDSS-I/II and SDSS-III/IV data (http://www.sdss.org/ and http://www.sdss3.org/). The National Radio Astronomy Observatory is a facility of the National Science Foundation operated under cooperative agreement by Associated Universities, Inc. This paper includes data gathered with the Apache Point Observatory 3.5 m telescope, which is owned and operated by the Astrophysical Research Consortium. 

This research has made use of software provided by the \chandra\ X-ray Center in the application packages CIAO, ChIPS, and Sherpa.

Funding for the Sloan Digital Sky Survey IV has been provided by the Alfred P. Sloan Foundation, the U.S. Department of Energy Office of Science, and the Participating Institutions. SDSS-IV acknowledges support and resources from the Center for High-Performance Computing at the University of Utah. The SDSS web site is www.sdss.org.

SDSS-IV is managed by the Astrophysical Research Consortium for the Participating Institutions of the SDSS Collaboration including the Brazilian Participation Group, the Carnegie Institution for Science, Carnegie Mellon University, the Chilean Participation Group, the French Participation Group, Harvard-Smithsonian Center for Astrophysics, Instituto de Astrof\'isica de Canarias, The Johns Hopkins University, Kavli Institute for the Physics and Mathematics of the Universe (IPMU)/University of Tokyo, Lawrence Berkeley National Laboratory, Leibniz Institut f\"ur Astrophysik Potsdam (AIP),  Max-Planck-Institut f\"ur Astronomie (MPIA Heidelberg), Max-Planck-Institut f\"ur Astrophysik (MPA Garching), Max-Planck-Institut f\"ur Extraterrestrische Physik (MPE), National Astronomical Observatories of China, New Mexico State University, New York University, University of Notre Dame, Observat\'ario Nacional/MCTI, The Ohio State University, Pennsylvania State University, Shanghai Astronomical Observatory, United Kingdom Participation Group,Universidad Nacional Aut\'onoma de M\'exico, University of Arizona, University of Colorado Boulder, University of Oxford, University of Portsmouth, University of Utah, University of Virginia, University of Washington, University of Wisconsin, Vanderbilt University, and Yale University.

{\it Facilities}: APO 3.5m (DIS), {\it CXO} (ACIS), {\it HST} (WFC3), Sloan, VLA.

\software{CASA \citep{McMullin2007}, CIAO \citep{Fruscione2006}, Sherpa \citep{Freeman2001,Doe2007}, ChiPS \citep{Germain2006}, pPXF \citep{Cappellari2004}, qsofit \citep{Guo2018,Shen2019}}

%%%%%%%%%%%%%%%%%%%% REFERENCES %%%%%%%%%%%%%%%%%%
%\bibliography{binaryrefs}
\bibliography{/Users/zeus/Documents/References/binaryrefs}

%\clearpage

\end{document}